\documentclass[12pt]{article}
\usepackage[hidelinks,colorlinks=true,linkcolor=blue,citecolor=blue,urlcolor=blue]{hyperref}
\usepackage[round,authoryear]{natbib}
\usepackage{graphicx}
\usepackage{booktabs}
\usepackage{amsmath} 
\usepackage{amsfonts}

\usepackage{times}
\makeatletter
\renewcommand\paragraph{\@startsection{paragraph}{4}{\z@}%
  {3.25ex \@plus 1ex \@minus .2ex}%
  {1.5ex \@plus .2ex}%
  {\normalfont\normalsize\bfseries}}
\makeatother

\newcounter{lastnote}

\title{Discovering Governing Spatial Interaction Equations in Dynamic Urban Systems}

\author
{\href{https://orcid.org/0009-0008-4416-5233}{Zhongfu Ma},$^{1}$ \href{https://orcid.org/0000-0002-3237-6032}{Di Zhu},$^{1\ast}$\\
\\
\normalsize{$^{1}$University of Minnesota, Twin Cities,}\\
\normalsize{Department of Geography, Environment, and Society, Minnesota, U.S.}\\
\\
\normalsize{$^\ast$To whom correspondence should be addressed; E-mail:  dizhu@umn.edu.}
}

\date{}

\begin{document} 

\baselineskip10pt


\maketitle

\begin{abstract}
Governing equations are fundamental for describing and predicting dynamic urban geographic systems.
Unlike physical systems guided by first principles, urban spatiotemporal phenomena emerge from coupled geographic processes that lack deterministic theoretical foundations, making the discovery of governing equations elusive and largely heuristic. 
Spatiotemporal dynamics in urban systems are often observed as sequential snapshots of spatial distribution, while the spatial interactions, or the flows, that cause such dynamics are often implied or unknown.
In this study, we propose a general differential equation formalism that decomposes urban dynamics into a spatial interaction process and a self-dynamic component. 
Building on this formalism, we introduce the Urban Discovery Framework (U-Discovery), which integrates hypothesis generation, neural fitting, and governing equation identification for discovering the governing equations of spatial interaction.
U-Discovery leverages Large Language Models and literature-based reasoning to propose differential equation candidates. 
Each candidate was calibrated from the observed spatiotemporal dynamics using a neural fitting method. 
The candidates are evaluated and ranked based on the fitting error and mathematical complexity. 
A synthetic experiment using simulated dynamics demonstrates that U-Discovery can identify the governing equation from a candidate set.
Empirical experiments examine real-world active population dynamics in Hennepin County, Minnesota. Drawing on both literature-derived scientific priors and broader knowledge from LLMs, U-Discovery proposes original candidate equations and identifies those that best reproduce observed population dynamics.
The discovered equations show an ability to recover the underlying mobility flows, while their varying performance across spatial scales suggests that different interaction mechanisms may dominate at different scales.

\end{abstract}

\newpage

\section{Introduction}

Urban dynamics have long been studied from a view of complex systems \citep{albeverio2008dynamics}, where the functional regions are linked by spatial interactions that are manifested by flows of people, goods, and information \citep{batty2013new, shiurban}. 
These flows redistribute people and resources, driving changes in urban systems over time.
In the era of geospatial big data, urban dynamics have been captured from diverse aspects and at increasingly fine spatiotemporal resolutions, revealed through patterns of socioeconomic activity \citep{zhu2017urban, xu2023urban}, emergency responses \citep{bagrow2011collective}, and epidemic spreading \citep{huang2020twitter,aguilar2022impact}.
A governing equation of spatial interaction mathematically characterizes the relationship between spatial flows and changes in urban states. 
Discovering such an equation can deepen our understanding of flow-driven urban dynamics and inform infrastructure optimization, resilient urban design, and public health management \citep{pappalardo2023future}.

The governing equation of spatial interaction in dynamic urban systems has been approached from diverse disciplines. 
Originating from the physical domain, the gravity model adapts Newton’s law of gravitation to human mobility \citep{zipf1946p}, characterizing movement flows as a function of the population of interacting locations and the distance between them.
Classic gravity models were initially developed to describe interactions between regions at a single point in time.
For urban dynamics, subsequent work incorporated gravity-based interactions into differential equations to model changes in urban systems over time, such as the Harris–Wilson model \citep{harris1978equilibrium} and reaction-diffusion models \citep{reia2022modeling,jin2023detecting}.
Beyond gravity-based models, interaction mechanisms have also been interpreted using analogies drawn from other domains. 
For instance, the urban metabolism model views the city as a living organism where flows transfer energy and materials to sustain urban operations \citep{conke2015urban}, while socio-hydrodynamic models \citep{seara2025sociohydrodynamics} integrate principles of fluid mechanics to explain residential dynamics by continuous flows of population migration.

The aforementioned studies provide general theoretical foundations and mathematical descriptions of spatial interaction mechanisms in dynamic urban systems.
However, for a particular observed urban dynamic, identifying the governing spatial interaction equation that best characterizes the observed dynamics remains a critical challenge.
On the one hand, unlike physical systems governed by established fundamental laws, urban systems lack first-principle knowledge to guide model development \citep{moroni2015complexity, lyu2025video}. 
Therefore, identifying plausible equation forms of spatial interaction has traditionally relied on domain intuition and iterative exploration, which is further compounded by inherent uncertainty and heterogeneity of human activity \citep{song2010limits}.
On the other hand, the dynamics of urban systems are typically captured as discrete snapshots of state, such as population distributions at subsequent time steps, while the interaction flows that drive these changes are often unobservable \citep{xu2025predicting, zhu2025gravity}.
Although data-driven discovery approaches have been increasingly applied to complex systems \citep{yu2025discover, guo2025sr}, the absence of observed spatial interactions limits their direct application to dynamic urban systems, where the effects of spatial interaction must be disentangled from other drivers of urban change.
This motivates a data-driven discovery paradigm grounded in a general differential-equation framework that explicitly separates the spatial interaction term from other drivers, enabling its inference from snapshot observations.

In this study, we introduce an Urban Discovery framework (U-Discovery) to reveal the unknown governing equation of spatial interaction in dynamic urban systems without requiring direct observations of interaction flows.
First, we embed the spatial interaction equation into a general differential-equation framework for urban dynamics.
The framework decomposes the state change of an urban system into two components: a spatial evolution term that uses spatial interaction equation to represent redistribution across locations, and a temporal dynamics term that accounts for periodic local dynamics, such as daily human rhythms.
To propose plausible spatial interaction equations, we leverage large language models (LLMs) to generate equation candidates grounded in scientific priors, which are extracted from interdisciplinary literature using Graph-based Retrieval-Augmented Generation (GraphRAG). 
In order to test various equation candidates on data, we develop a neural fitting method, Urban Differential Equation Network (UrbanDE-Net), which enables efficient and stable parameter estimation for equation candidates.
The potential governing equation can be identified by ranking candidates using a joint criterion of fitting performance and equation complexity.
We first validate U-Discovery through synthetic experiments, demonstrating that the framework can successfully identify the ground-truth interaction equation from a candidate set. 
We then apply U-Discovery to real-world human mobility data to identify the spatial interaction equation underlying active population dynamics in Hennepin County, Minnesota, U.S.
A newly discovered spatial interaction equation outperforms traditional gravity-based formulations and demonstrates its ability to reproduce both the temporal dynamics of active population and the mobility flows driving those dynamics.
U-Discovery provides a data-driven paradigm for uncovering governing equations of spatial interaction directly from spatiotemporal observations of urban systems, bridging the gap between complex-system theory and empirical urban dynamics.

The paper is organized as follows: 
In Section \ref{sec: Related work}, we review the related work on spatial interaction modeling for dynamic urban systems and data-driven equation discovery. 
In Section \ref{sec: Method}, we formalize the problem and introduce the U-Discovery framework, including hypothesis generation via GraphRAG, candidate fitting using UrbanDE-Net, and the evaluation strategy for governing equation identification.  
Section \ref{sec: syn} validates our framework through synthetic experiments to demonstrate its capability in recovering the ground-truth equation. 
Section \ref{sec: real exp} presents the empirical application of U-Discovery using real-world human activity data in Hennepin County, Minnesota. 
Section \ref{sec: disscussion} explores the scale dependence in equation discovery, and discusses limitations and future research directions
Finally, we conclude this study in Section \ref{sec: conclusion}.

\section{Related work} \label{sec: Related work}

\subsection{Spatial interaction in dynamic urban systems}

Spatial interactions refer to the processes through which geospatial entities influence one another via the movement of population, information, goods, and capital across space \citep{Fotheringham2001, Roy2003}. In dynamic urban systems, these interactions manifest as time-varying flows that continuously reshape the spatial distribution of activities and functions. 
A substantial body of research has examined spatial interaction from multiple perspectives. 
First, empirical studies leverage network analysis and community detection techniques to uncover the spatiotemporal organization of interaction patterns \citep{jia2022dynamical,liu2022revealing,zhang2024mining}, revealing functional regions \citep{tao2019re, zhu2020understanding, zhu2026uncovering, xiong2026sensing}, and mobility communities \citep{zhong2014detecting,ma2025collective, su2022classification}. 
These studies characterize how interaction intensities evolve and how collective flow patterns reflect the mesoscopic structure of cities. 
Second, predictive research focuses on estimating or forecasting spatial interaction flows. 
With the availability of large-scale mobility data, machine learning and deep learning models have been increasingly adopted to infer origin–destination flows and capture nonlinear dependencies between spatial attributes and interaction intensity \citep{rong2021inferring, zhu2025gravity, wang2025context}. 
Such approaches enhance predictive accuracy but often provide limited interpretability regarding the mechanisms of spatial interaction.

To move beyond pattern description and black-box prediction toward explanatory understanding, a long tradition of research has developed explicit mathematical models that formalize spatial interaction as a function of place attributes and spatial separation.
Notable contributions include the gravity model \citep{zipf1946p}, the intervening opportunity model \citep{stouffer1940intervening}, and the radiation model \citep{simini2012universal}. 
These classic spatial interaction models are developed based on a static spatial configuration. 
To directly extend them to dynamic scenarios, efforts on the panel gravity model \citep{cameron2019estimation, pu2019spatial} extend the gravity model to account for the spatial interaction mechanism in sequential snapshots of spatial distribution. 
By integrating the spatial interaction mechanisms in dynamic models, researchers have adopted the models from a wide range of domains to explain the dynamics of the urban system.
For example, interaction-based mechanisms are embedded within cellular automata and agent-based models to simulate urban expansion, spatial diffusion, and regional transformation processes \citep{batty2007cities, torrens2001cellular}. 
Also, physical models and spatial economics models have embraced spatial interaction models to explain how local interactions can produce large-scale urban structural change and social segregation over time, such as the Harris–Wilson model \citep{harris1978equilibrium}, the reaction-diffusion model \citep{reia2022modeling,jin2023detecting}, and the sociohydrodynamic model \citep{seara2025sociohydrodynamics}.

\subsection{Data-driven discovery of governing equations}

Early work on data-driven equation discovery sought to recover compact, interpretable governing equations directly from observations by searching over symbolic forms. 
\cite{bongard2007automated} pioneered automated symbolic identification for nonlinear coupled dynamical systems from time-series data, demonstrating that governing equations can be recovered without fully specifying the model form. 
Subsequent studies improved both the search principles and the robustness of symbolic discovery; for example, \cite{schmidt2009distilling} emphasized criteria for identifying nontrivial laws and distilling free-form natural laws from experimental data. 
\cite{rudy2017data} introduced a sparse regression method to discover the governing partial differential equation that most accurately represents the data without searching through a large body of possible candidate models. 
Nevertheless, large search spaces, noise sensitivity, and the difficulty of extending discovery to high-dimensional interacting systems remain persistent challenges, motivating hybrid strategies that combine neural representation learning with symbolic regression. 
Recent neuro-symbolic approaches \citep{gao2022autonomous,yu2025discover} leverage deep models to guide symbolic search and uncover network dynamics at scale, improving efficiency and enabling discovery beyond small systems.

These methodological advances have also been applied to social, environmental, and Earth-system processes where first principle  are incomplete, and domain mechanisms are debated. 
In land-change modeling, \cite{manson2007agent} demonstrated symbolic regression in the context of agent-based deforestation models, illustrating how interpretable rules can be inferred for complex human–environment systems. 
For human mobility, \cite{guo2025distilling} used symbolic regression to systematically rediscover classic mobility model (e.g., gravity-like distance decay) and propose novel equation forms, while \cite{cabanas2025human} showed that Bayesian symbolic regression can yield gravity-like analytical forms that achieve competitive predictive performance with complex machine-learning models. 
From a broader geoscience perspective, \cite{song2024towards} reviewed data-driven equation discovery as a scientific-AI pathway and summarized opportunities and challenges across Earth-system applications. 
More recently, LLM-based discovery has emerged to incorporate scientific priors and steer the generation of equation candidates \citep{shojaee2024llm,wang2025drsr}, and retrieval-augmented generation has been explored to support incremental discovery and more complex analytical expressions \citep{guo2025sr}. 
In parallel, urban-focused LLM efforts such as \cite{li2024urbangpt} and \cite{liang2024exploring} have shown the promise of foundation models for prediction tasks, such as traffic flow and mobility under special events, suggesting a timely opportunity to incorporate LLMs in discovering interpretable equations using spatial big data.

\section{U-Discovery}\label{sec: Method}
\subsection{Definition and problem statement}
\textbf{Dynamic urban system}: 
We conceptualize the dynamics of an urban region, such as a metropolitan area, city, or county, as a dynamic spatial network consisting of $n$ spatial units as nodes.
At time $t$, the urban system is represented as a directed graph $G_t=(P_t, T_t, D)$, characterized by two observed properties, the time-varying node states $P_t$ and the static spatial structure $D$, and one unobserved property, the spatial interactions $T_t$. 
\begin{itemize}
    \item \textbf{Node states}: Each node $i$ is characterized by an observable, time-varying state $p_{i,t}$, such as the active population considered in this study.
    The states of all nodes at time $t$ form a system snapshot, denoted as $P_t=\{p_{1,t}, p_{2,t},...,p_{n,t}\}$.
    To characterize observed dynamics, we defined $\frac{d p_{i,t}}{dt}$ as the instantaneous rate of change of the state attribute at node $i$. 
    Although our empirical observations are available at discrete time intervals, we assume that urban dynamics are governed by a continuous-time process. 
    In practice, the time derivative can be approximated by finite differences between consecutive snapshots, i.e., $\frac{d p_{i,t}}{dt}\approx\frac{p_{i,t+\Delta t}-p_{i,t}}{\Delta t}$, where $\Delta t$ is the time interval between two consecutive snapshots.
    \item \textbf{Spatial structure}: The observed edge property in $G_t$ is represented as $d_{ij}$, which denotes the spatial separation from spatial unit $i$ to spatial unit $j$. 
    The matrix $D=\{d_{ij}|i\neq j\}$ therefore characterizes the spatial structure through which interactions may occur.
    Depending on the observed representation of spatial structure, $D$ may be symmetric, such as Euclidean distance, or asymmetric, such as directional travel time or network cost.
    \item \textbf{Spatial interactions}: Spatial interactions constitute the unobserved edge property of the urban system and represent the spatial redistribution process underlying the observed node-state dynamics.
    They are represented as $T_t=\{T_{ij,t}|i\neq j\}$, where $T_{ij,t}$ denotes the directed interaction, or flow, from spatial unit $i$ to spatial unit $j$.
    Because the interaction topology is assumed to be unknown a priori, every ordered pair of distinct nodes is considered to have a potential interaction. 
    We therefore define $G_t$ as a complete directed graph, in which $T_{ij,t}$ and $T_{ji,t}$ are treated as different interactions with opposite direction between nodes.
\end{itemize}

\textbf{Governing differential equation}: 
We assume that there is a time-invariant mechanism governing the transition between each subsequent time step, e.g., from $P_t$ to $P_{t+1}$, based on the spatial structure of the urban system $D$. 
Drawing on the widely used analytical framework of network dynamics \citep{barzel2013universality, gao2022autonomous, yu2025discover}, we represent urban dynamics using a general differential equation that decomposes the rate of change of node $i$, $\frac{d p_{i,t}}{dt}$, into three components:
\begin{equation}
     \frac{d p_{i,t}}{dt} = f_{se}(P_{t}, D; \theta_{se})+f_{td}(t; \theta_{td})+\varepsilon_{i,t},
\end{equation}
These components are defined as follows: 
\begin{itemize}
    \item \textbf{Spatial evolution term $f_{se}(\cdot)$}: This term captures the aggregate effect of spatial interactions on node $i$. Each directed interaction is generated by an unknown spatial interaction equation:
    \begin{equation}
        \qquad T_{ij,t}=SI(z_{ij,t}; \theta_{se}), \qquad z_{ij,t}\in Z(P_{t}, D),
    \end{equation}
    where $SI$ denotes the time-invariant spatial interaction equation and $\theta_{se}$ denotes its parameter set.
    $z_{ij,t}$ denotes the variables and derived quantities used by the equation, while $Z(P_t, D)$ represents the available information that can be obtained from the current system state $P_t$ and spatial structure $D$.
    Given the $SI$, $f_{se}(\cdot)$ is defined as the sum of the net directed interactions between unit $i$ and all other spatial units. Its contribution to the state change over an infinitesimal interval $dt$ is therefore
    \begin{equation}
        dp_{i,t}^{se}=f_{se}(P_{t}, D; \theta_{se})dt=\sum_{j=1}^n (T_{ji,t}-T_{ij,t})dt,
    \end{equation}
    where $T_{ji,t}$ and $T_{ij,t}$ denote the inflow from node $j$ to node $i$ and outflow from node $i$ to node $j$, respectively.
    
    \item \textbf{Temporal dynamics term $f_{td}(\cdot)$}: This term accounts for periodic patterns inherently influencing the system, such as daily rhythms, where $\theta_{td}$ is the parameter set of $f_{td}(\cdot)$. 
    The temporal contribution to the state change is expressed as $dp_{i,t}^{td}=f_{td}(\cdot)d t$, which refers to the change in periodicity for node $i$ within a time step. 
    
    \item \textbf{Noise term $\varepsilon_{i,t}$}: This term represents stochastic fluctuations, measurement errors, and other unmodeled effects. 
    It is included in the general governing equation but is not explicitly modeled during equation discovery.
\end{itemize}

The governing differential equation provides an analytical framework that decomposes the state-by-state dynamics of the urban system into spatial interactions among places and the temporal dynamics intrinsic to each place. 
Building upon this formalism, the objective of this study is to discover the spatial interaction equation that best explains the observed urban dynamics. 
Formally, a set of plausible equation candidates, $SI=\{SI^1, SI^2,..., SI^m\}$, is proposed to construct the spatial evolution term, representing different mechanisms of spatial interaction (e.g., gravity-based flows, diffusion, or radiation).
These candidates are then embedded into the full governing differential equation alongside the temporal dynamics term, and fitted to the observed system dynamics, given a sequence of observed system snapshots $P=\{P_1, P_2, ...P_m\}$ and the underlying spatial structure $D$
By evaluating the performance of each candidate model regarding the accuracy of reproducing the observed dynamics, we can identify the best-fitting spatial interaction equation from our candidate set to approximate the governing equation of the dynamic urban system.

\subsection{Urban Discovery Framework}\label{sec: Framework}
\begin{figure}[!ht]
    \centering
    \includegraphics[width=\linewidth]{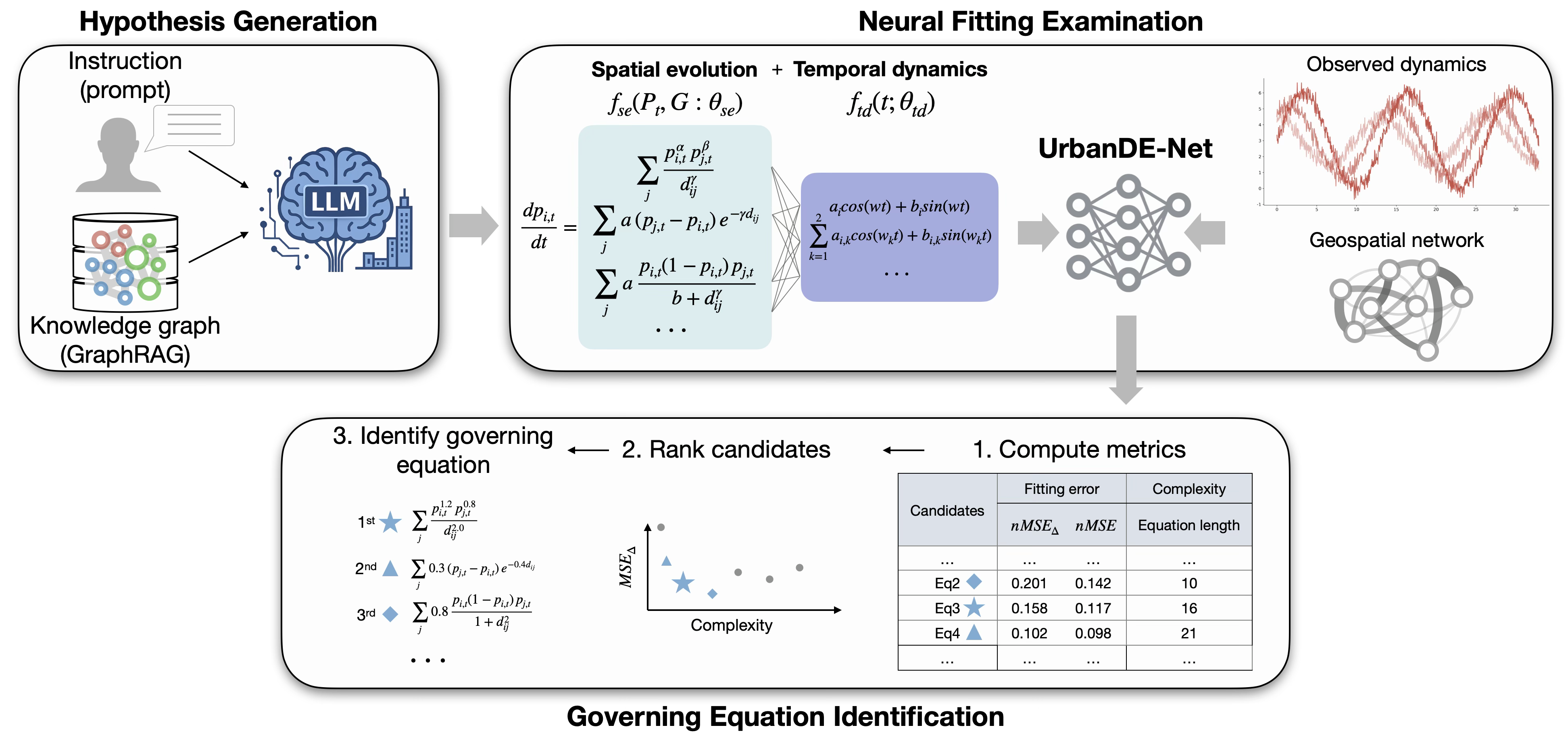}
    \caption{The Urban Discovery framework (U-Discovery). 
    In Hypothesis Generation (HG), GraphRAG is utilized to construct a knowledge graph from existing literature, extracting scientific priors as context for a Large Language Model (LLM) to propose equation candidates for spatial interaction and form spatial evolution term.
    In Neural Fitting Examination (NFE), the Urban Differential Equation Network (UrbanDE-Net) evaluates these candidates against observed dynamics by fitting the differential terms.
    In Governing Equation Identification (GEI), candidate equations are ranked based on fitting error and complexity. The equation that optimally balances these two metrics is identified as the potential governing equation from the candidate set.}
    \label{fig: Framework}
\end{figure}

In this section, we introduce the architecture of our proposed framework, the Urban Discovery (U-Discovery). As shown in Figure \ref{fig: Framework}, U-Discovery consists of three steps:
\begin{itemize}
\item \textbf{Hypothesis Generation (HG):} Utilizing a Large Language Model (LLM) augmented by a domain-specific knowledge base (GraphRAG), this module formulates a diverse pool of candidate spatial interaction equations to form the spatial evolution term, which explicitly combine existing scientific priors.
\item \textbf{Neural Fitting Examination (NFE):} The generated equation candidates, combined with temporal dynamics term, are integrated into Urban Differential Equation Network (UrbanDE-Net), a graph-based neural fitting method to evaluate their validity against observed spatiotemporal urban dynamics. 
\item \textbf{Governing Equation Identification (GEI):} 
Finally, the framework systematically ranks the evaluated candidates to identify the optimal governing equation. 
Collectively comparing fitting error and complexity, this module identifies the mathematical formulation that offers the best balance between predictive accuracy and form complexity.
\end{itemize}

\subsubsection{Hypothesis Generation via Large Language Models and GraphRAG}\label{sec: candidates}
\begin{figure}[!ht]
    \centering
    \includegraphics[width=0.9\linewidth]{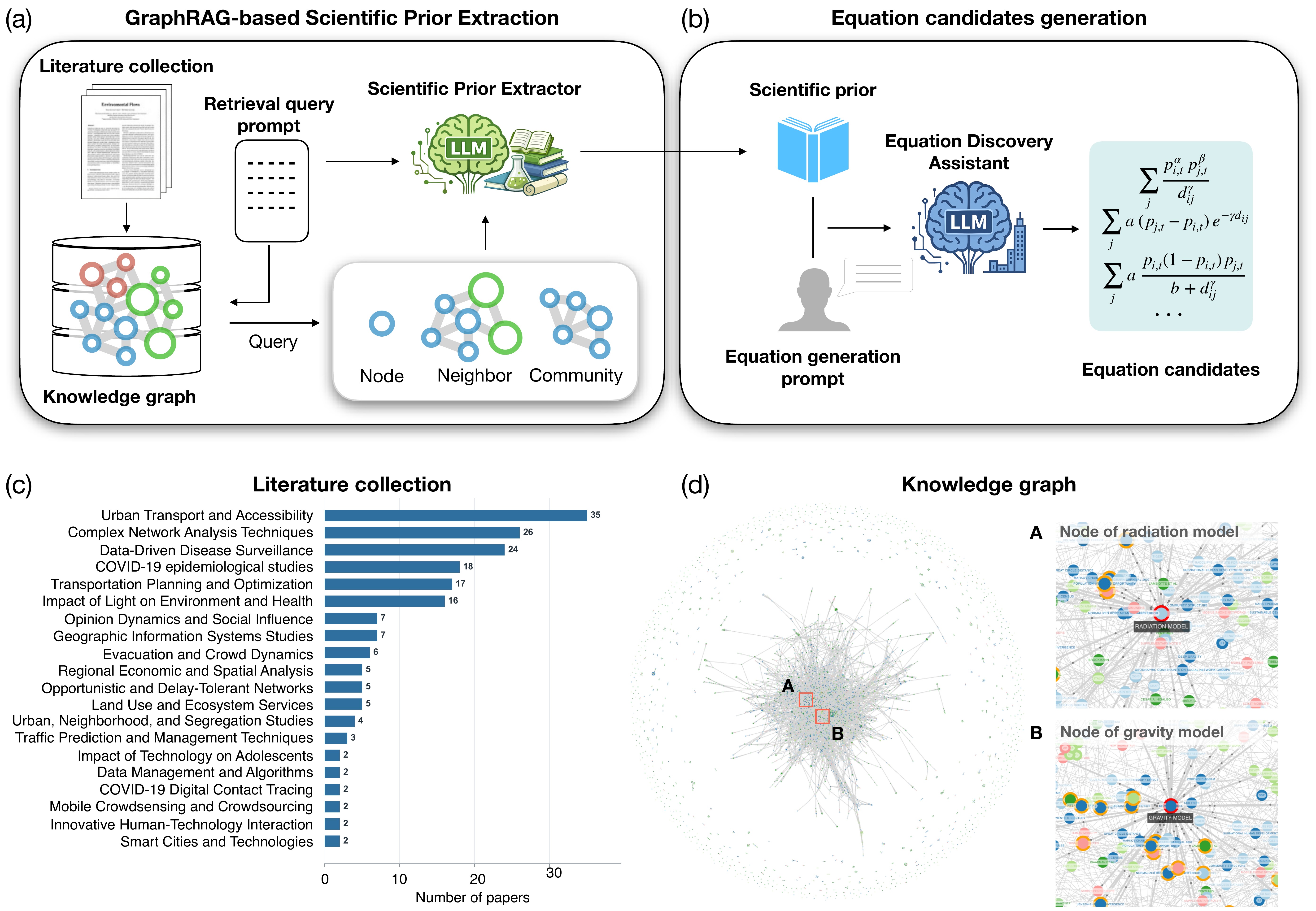}
    \caption{Workflow for extracting scientific priors and generating equation candidates via two LLM-based agents, together with the literature collection and resulting knowledge graph.
    (a) The Extractor retrieves knowledge from a literature-based knowledge graph via GraphRAG. 
    (b) The Assistant synthesizes the extracted priors to propose mathematical equations for the spatial interaction and form spatial evolution terms.
    Scientific Prior Extractor and Equation Discovery Assistant are two LLM-based agents used in this workflow. 
    (c) Topic frequencies in the literature collection. 
    (d) The global knowledge graph and enlarged local neighborhoods of the concept nodes representing the radiation model (A) and gravity model (B).
    }
    \label{fig: HG_prompt}
\end{figure}

To propose candidates of the spatial interaction equation, a large body of research on spatial interaction modeling and urban systems has offered diverse assumptions and functional forms to explain how spatial interactions emerge from the uneven distribution of resources, spatial constraints, and distance frictions (see Section \ref{sec: Related work}). 
At the same time, studies in a broader domain, such as network science and statistical physics, have developed transferable laws that can inspire the proposal of new candidate terms.
However, these insights are dispersed across disciplines, expressed in different terminology, and embedded in descriptive text instead of mathematical formulas, making them difficult to incorporate when proposing interpretable spatial interaction equations.
To address this challenge, we introduce an LLM-based method: first, we adopt a Graph-based Retrieval Augmented Generation (GraphRAG) \citep{edge2024local} to construct a knowledge graph of literature for retrieval scientific priors (Figure \ref{fig: HG_prompt} (a)); second, we leverage LLMs to assemble the retrieved knowledge as context and use it to propose equation candidates of spatial interaction and form the spatial evolution term (Figure \ref{fig: HG_prompt} (b)).

The extraction of scientific priors starts from converting the literature collection into a knowledge graph, which enables the systematic retrieval of dispersed concepts and their complex relationships.
We first partition the literature collection into discrete text units, which serve as the basic units for knowledge graph extraction and provenance tracking.
For each text unit $u_i$, an LLM extracts entities and relationships as nodes and edges, respectively, to form a local subgraph. 
These local subgraphs are then integrated into a global knowledge graph by taking the union of all extracted nodes and edges.
Subsequently, we apply the Leiden algorithm \citep{traag2019louvain} to detect a set of hierarchical communities in the global knowledge graph. 
Finally, the LLM generates textual summaries for all of the entities, relationships, and communities.
These summaries, alongside the raw text units, are mapped into the embedding space to enable semantic retrieval.
In Figure \ref{fig: HG_prompt} (c) and (d), we provide an illustration of the literature collection and the knowledge graph constructed using the GraphRAG pipeline.
Figure \ref{fig: HG_prompt} (c) summarizes cross-disciplinary coverage of the literature collection, with 108 high-impact literature across 20 topics obtained from the OpenAlex\footnote{https://openalex.org/} database, spanning transportation, regional economics, disease surveillance, and other urban- and network-related domains. 
After extracting textual data from these publications, we utilized OpenAI's GPT-5 models to extract entities and relationships, resulting in a comprehensive knowledge graph comprising 14,769 nodes and 32,331 edges (see Figure \ref{fig: HG_prompt}(d)), and then embedded the graph using the text-embedding-3-small model \citep{openai2024textembedding3small}. 
Within the graph, the gravity and radiation models appear as highly connected concept nodes, illustrating the GraphRAG pipeline’s ability to identify and contextualize established spatial interaction mechanisms.

To formulate the scientific prior, we adapt the Local Search mechanism from the GraphRAG framework \citep{edge2024local}. 
Given a retrieval query prompt (see Figure \ref{fig: prompt}(a) in Appendix \ref{app: prompts}) that describes the requirement of scientific priors, we compute its embedding and retrieve top-$k$ semantically related entities by comparing the cosine similarity between their embeddings. 
Beyond entity-level search, we leverage the topological structure of the knowledge graph to expand our retrieval. 
By traversing from the top-$k$ entities to their neighbor, parent communities, and originating text units, we assemble a structured text as an enriched and comprehensive context set for scientific prior extraction.
An LLM-based agent, Scientific Prior Extractor, takes the context set as input and distills it into explicit building blocks for the spatial evolution terms.
Following the retrieval query prompt, these blocks include statements of the spatial interaction mechanisms, corresponding quotations of the raw statements, mathematical formulas (such as distance decay functions), and a library of symbols and parameters. 
Together, these form the scientific priors for proposing equation candidates.

Following extraction, the Equation Discovery Assistant takes the scientific priors and proposes spatial interaction equation candidates to form the spatial evolution term $f_{se}$ (Figure \ref{fig: HG_prompt} (b)), guided by the equation generation prompt (see Figure \ref{fig: prompt}(b) in Appendix \ref{app: prompts}).
To balance consistency of extracted priors with exploratory generation, we implement two distinct operational modes for this prompt: No Invention Mode restricts the Equation Discovery Assistant to scientific priors extracted from the literature when proposing spatial evolution terms.
Partial Invention Mode allows the Equation Discovery Assistant to incorporate knowledge beyond the extracted scientific priors when proposing spatial evolution terms.
Note that these two modes only constrain the knowledge availability for equation proposal but do not change the LLM's flexibility in translating and synthesizing the permitted knowledge into specific functional forms.
In the synthetic experiment, we apply the no invention mode to test whether the LLM agent can propose equation candidates strictly grounded in the retrieved scientific priors.
In the empirical study, we switch to the partial invention mode to explore its capacity to move beyond existing scientific priors and propose novel equation forms.
Finally, the Equation Discovery Assistant enforces a highly structured output format, which explicitly details equation candidates, parameter definitions, and supporting evidence. 
This structured text is designed to be easily converted into programming code, allowing for seamless integration with our subsequent neural fitting method. 
Furthermore, by explicitly linking the generated terms back to the extracted evidence, this framework ensures that the discovered equations remain highly interpretable and can be reliably traced back to their foundational literature.

\subsubsection{Neural Fitting Examination using the Urban Differential Equation Network}\label{sec: fitting}
Validating candidate models across diverse domains requires a flexible, equation-agnostic fitting method. 
Recent neural network-based calibration methods \citep{gaskin2023neural} estimate dynamical equation parameters more efficiently and accurately than classical approaches like Markov chain Monte Carlo \citep{stuart2010inverse}. 
This offers a general computational pathway to test and compare mechanistic hypotheses on large-scale data. 
Motivated by this, we extend the neural fitting method in \cite{ma2025neural} and introduce the Urban Differential Equation Network (UrbanDE-Net) to fit governing differential equations to observed urban dynamics.

\begin{figure}[!ht]
    \centering
    \includegraphics[width=0.9\linewidth]{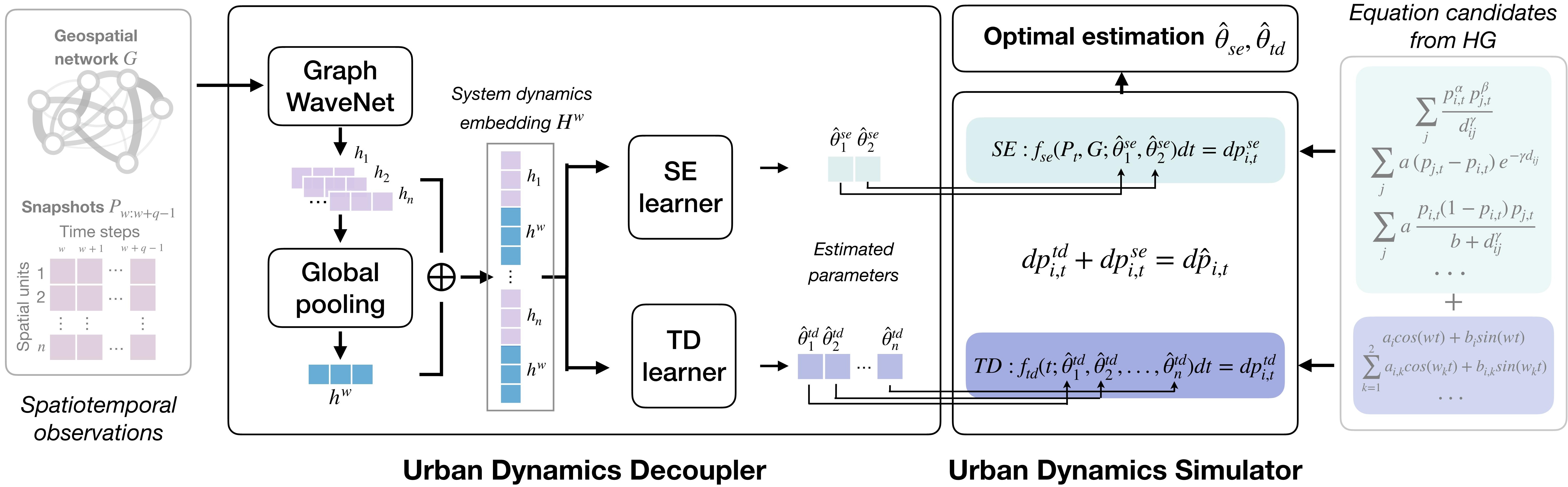}
    \caption{Model architecture of the Urban Differential Equation Network (UrbanDE-Net). 
    The Urban Dynamics Decoupler takes spatial structure $D$ and a time window of snapshots $P_{w:w+q-1}$ as input to first generate the system dynamics embedding $H^w$ as a deep representation of the underlying mechanisms. 
    The SE and TD learners then decouple this embedding to estimate the parameters, $\hat{\theta}^{se}$ and $\hat{\theta}^{td}$, for a set of equation candidates. 
    Finally, the Urban Dynamics Simulator integrates the equation candidates proposed by the HG to reconstruct the snapshots. 
    The reconstruction error is then backpropagated to update the weights in the Decoupler, optimizing the parameter estimation process.
    }
    \label{fig: UrbanDE-Net}
\end{figure}

UrbanDE-Net consists of two key components (illustrated in Figure \ref{fig: UrbanDE-Net}):
\begin{itemize}
\item \textbf{Urban Dynamics Decoupler}: This module employs a Graph WaveNet encoder to extract multi-scale spatial and temporal features from short observation windows, utilizing parallel MLP heads to decode these representations into interpretable parameters for separate spatial evolution and temporal dynamics components.
\item \textbf{Urban Dynamics Simulator}: This module functions as a numerical solver that reconstructs urban dynamics based on estimated parameters. The resulting reconstruction error, constrained by both first- and second-order temporal derivatives, is backpropagated to iteratively optimize the Decoupler for more accurate parameter identification.
\end{itemize}

\paragraph*{Urban Dynamics Decoupler}
Urban Dynamics Decoupler adopts a Graph WaveNet (GWN) encoder \citep{wu2019graph} to learn a deep representation of spatiotemporal urban dynamics from a short observation window, and then two MLP heads: Spatial Evolution (SE) Learner and Temporal Dynamics (TD) Learner, decompose this representation into parameters of candidate differential equation components.
GWN encoder captures long-range temporal dependencies via stacked dilated causal convolutions and models spatial dependencies through a diffusion-style graph convolution augmented by a self-adaptive adjacency matrix learned from node embeddings.

The GWN encoder takes a window $P_{w:w+q-1}=\{P_w, P_{w+1},...,P_{w+q-1}\}$ as input, where $w$ is the start time step of the window, $q$ refers to the window size, and each snapshot $P_t=\{p_{1, t}, p_{2, t}, ...,p_{n, t}\}$ records the node states at time $t$. 
To preprocess the data as model input, $P_{w:w+q-1}$ is arranged as $X\in \mathbb{R}^{n\times c\times q}$, where the feature dimension $c=1$, and projected to the hidden space by a $1\times 1$ convolution:
\begin{equation}
    H^{(0)}=Conv_{1\times 1}(X).
\end{equation}
GWN encoder stacks $K$ spatio-temporal layers. 
Each layer $\ell$ applies a gated dilated causal convolution along the temporal dimension to extract temporal patterns at different time scales, while preserving causality:
\begin{equation}
\tilde{H}^{(\ell)}=\tanh\!\left(W_f^{(\ell)} \star H^{(\ell-1)}\right)\odot
\sigma\!\left(W_g^{(\ell)} \star H^{(\ell-1)}\right),
\end{equation}
where $H^{(\ell-1)}$ denotes the layer input feature map, $W_f^{(\ell)}$ and $W_g^{(\ell)}$ are learnable convolution kernels, and $\star$ denotes a dilated causal convolution along the temporal axis. 
The nonlinearities $\tanh(\cdot)$ and $\sigma(\cdot)$ (sigmoid) produce the candidate signal and a gate, and $\odot$ is the element-wise product that adaptively controls information flow.
This design enables a large receptive field with only a few layers, allowing the encoder to summarize the temporal mechanism from both short-term fluctuations and longer-range trends within the window.

Given the temporal features $\tilde{H}^{(\ell)}$, spatial dependency is modeled by a diffusion-style graph convolution, by which node information is repeatedly propagated along the graph structure so that each node can aggregate signals from neighborhoods (Eq. \ref{eq:diffu}), naturally matching the notion of spatial evolution driven by interactions among locations.
\begin{equation}\label{eq:diffu}
A\tilde{H}^{(\ell)}=
\sum_{j=1}^{n} A_{ij}\,\tilde{H}^{(\ell)},
\end{equation}
where $A\in \mathbb{R}^{n\times n}$ is the adjacency matrix of the graph.
Besides using $A_{dis}$ for distance-weighted graph structure, we also incorporate learned adaptive adjacency
\[
A_{adp}=\operatorname{softmax}\!\left(\operatorname{ReLU}(E_1E_2^\top)\right),
\]
where $E_1, E_2 \in \mathbb{R}^{n\times c}$ are learnable node embeddings.  models the interaction mechanism when the distance is not the only factor in a spatial evolution term. 
The diffused features are concatenated and projected back to the hidden dimension $d$ using a transformation $\Phi(\cdot)$:
\begin{equation}
H^{(\ell)} = \Phi \left( \left[ \tilde{H}^{(\ell)} \parallel A_{dis} \tilde{H}^{(\ell)} \parallel A_{adp} \tilde{H}^{(\ell)} \right] \right) \in \mathbb{R}^{n \times d \times q},
\end{equation}
where $\parallel$ denotes the concatenation operator along the feature dimension.

To summarize the dynamics within the observation window, the encoder first extracts node-level embeddings $h_i$ by applying temporal mean pooling and a linear projection across the final hidden states $H^{(K)}$. These node-specific features are then aggregated into a single graph-level embedding $h^w$ via a global attention pooling mechanism:
\begin{equation}
h_i = \text{Linear} \left( \frac{1}{q} \sum_{t=1}^{q} H_{i, d, t}^{(K)} \right), \quad h^w = \sum_{i=1}^{n} \alpha_i h_i
\end{equation}
where $h_i, h^w \in \mathbb{R}^d$, and $\alpha_i$ is learnable weights. This dual-level representation allows the Decoupler to capture both localized fluctuations and the collective evolution of the entire urban system.

To estimate the parameters for each equation candidate, we first fuse local (node-level) and global (graph-level) representations of the dynamics. Specifically, given node embeddings $h_i$ and a global embedding $h^w$ summarizing all the nodes of an input window, we broadcast $h^w$ to all nodes and concatenate it with each $h_i$:
\begin{equation}
h_i^{w} = [h_i \Vert h^{w}] \in \mathbb{R}^{2d},
\quad
H^{w} = \{h_i^{w}\}_{i=1}^{n}.
\end{equation}
The fused representations $H^{w}$ are then passed to a decoupled parameter learner implemented as two parallel MLP heads. The Spatial Evolution head $MLP_{\mathrm{se}}$ estimates the parameters governing the spatial evolution term, while the Temporal Dynamics head $MLP_{\mathrm{td}}$ estimates the parameters of the temporal component:
\begin{equation}
\hat\theta_{\mathrm{se}} = MLP_{\mathrm{se}}(H^{w}),
\quad
\hat\theta_{\mathrm{td}} = MLP_{\mathrm{td}}(H^{w}).
\end{equation}
UrbanDE-Net supports two fitting modes, global and local, to accommodate spatial homogeneity and location-specific heterogeneity while retaining a common functional form for a given candidate equation across the urban system.
Under global fitting, each parameter learner estimates a single parameter set shared across all spatial units, $\hat{\theta}_{\mathrm{se}}^{w}$ or $\hat{\theta}_{\mathrm{td}}^{w}$. 
This mode represents spatial homogeneity by assuming that the same set of equation coefficients governs all spatial units.
Under local fitting, the learner output is expanded along the node dimension to unit-specific parameter sets, 
$\{\hat{\theta}_{\mathrm{se},i}^{w}\}_{i=1}^{n}$ or $\{\hat{\theta}_{\mathrm{td},i}^{w}\}_{i=1}^{n}$. 
In contrast, local fitting represents spatial heterogeneity by allowing the equation coefficients to vary across spatial units while preserving the same functional form \citep{kordi2016spatially}.
Given that $H^{w}$ combines node-level representations with a network-level representation computed from observation window $w$, the same encoder architecture can support both fitting modes.

\paragraph*{Urban Dynamics Simulator}

To assess the estimated parameters for a given equation candidate, we compare the ground truth observations with the reconstructed dynamics. Under the assumption of a first-order Markov process, each state is derived directly from the preceding state based on the governing equation and the spatial structure $D$.
Given an estimated parameter set $\hat{\theta}_{se}$ and $\hat{\theta}_{td}$ from the Decoupler, we employ the Urban Dynamics Simulator as a numerical solver that maps the current observed state and graph context to the predicted instantaneous rate of change:
\begin{equation}
\frac{d \hat{p}_{i,t}}{dt} = f_{se}(P_{t}, D; \hat \theta_{se}) + f_{td}(t; \hat \theta_{td}).
\end{equation}
The Simulator is coupled with the Decoupler, which estimates window-specific parameters $\hat{\theta}_{se}$ and $\hat{\theta}_{td}$ from a sliding window of snapshots.
For each time step $t$ inside window $P_{w:w+q-1}$, we apply the Simulator to produce this one-step dynamic signal without modeling the noise and compute the window loss $\mathcal{L}_w$, which is then backpropagated to update the weights in the Decoupler.
The window loss $\mathcal{L}_w$ incorporates the Mean Squared Error (MSE) of both the first-order and second-order temporal derivatives on node states:
\begin{equation}
\mathcal{L}_w(\hat{\theta}_{se},\hat{\theta}_{td})
= \mathrm{MSE}\left(\frac{dp}{dt}, \frac{d\hat{p}}{dt}\right)
+ \lambda\ \mathrm{MSE}\left(\frac{d^2p}{dt^2}, \frac{d^2\hat{p}}{dt^2}\right),
\end{equation}
where $\lambda$ is a hyperparameter controlling the strength of the second-order constraint.
In discrete time, we approximate the second-order derivative using finite differences: $\frac{d^2\hat{p}_{t}}{dt^2} \approx \frac{\Delta\hat{p}_{t+1} - \Delta\hat{p}_{t}}{\Delta t^2}$, where $\Delta\hat{p}_{i, t}=\frac{d \hat{p}_{i,t}}{dt}\Delta t$.
While the first-order term encourages accurate reconstruction of the instantaneous change, the second-order term further constrains how the change itself varies over time, improving stability under noisy observations.

Training follows a three-step loop across epochs:
\begin{enumerate}
    \item \textbf{Parameter estimation:} Use the Urban Dynamics Decoupler to estimate window-specific parameters from a given time window.
    \item \textbf{One-step simulation:} Simulate the one-step dynamics between consecutive snapshots using the equation candidate and the estimated parameters.
    \item \textbf{Optimization:} Compute the loss and backpropagate gradients to update the weights in the Decoupler, thereby improving parameter estimation.
\end{enumerate}
After scanning all windows in an epoch, we evaluate the model on a validation set to obtain $\mathcal{L}_{\mathrm{val}}$.
If $\mathcal{L}_{\mathrm{val}}$ improves upon the best validation loss so far, we checkpoint the corresponding model weights and parameter estimates.

\subsubsection{Governing Equation Identification}\label{sec: evaluation}

To identify the potential governing equation from candidates, we evaluate from two dimensions: accuracy in reconstructing the observation and equation complexity reflected by equation length.
We evaluate the accuracy using two complementary error metrics computed on the absolute states and the step-to-step changes. 
Given the sequence of ground truth observations $P_t$ over $t=1,\ldots, T$, we generate the reconstructed snapshots $\hat{P}_{t+1}$ using the one-step predictions from the Urban Dynamics Simulator, such that $\hat{P}_{t+1} = P_{t} + \Delta\hat{P}_{t}$, where $\Delta \hat P_t=\{\frac{d \hat{p}_{1,t}}{dt}\Delta t, \frac{d \hat{p}_{2,t}}{dt}\Delta t, ...,\frac{d \hat{p}_{n,t}}{dt}\Delta t\}$. 
The state-level error is then measured by $\mathrm{MSE}(P_t,\hat{P}_t)$, which quantifies overall reconstruction accuracy. More importantly, because the governing equation is intended to explain the transition mechanism between snapshots, we emphasize the error in the step-to-step:
\begin{equation}\label{eq: mse_delta}
\mathrm{MSE}_{\Delta}=\frac{1}{T-1}\sum_{t=1}^{T-1}
\left\lVert
(\hat{P}_{t+1}-\hat P_t) - (P_{t+1}-P_t)
\right\rVert_2^2.
\end{equation}
Unlike standard $\mathrm{MSE}$, $\mathrm{MSE}_{\Delta}$ directly evaluates whether an equation candidate correctly reproduces both the trend and magnitude of the dynamics, making it better aligned with our goal of identifying the interaction-driven evolution term from panel observations. 
To quantify how much the spatial evolution term contributes to the overall accuracy, we introduce normalized metrics relative to a temporal-only baseline. 
Specifically, we generate a baseline trajectory $\hat{P}^{td}_t$ by fitting a governing equation only with the temporal dynamics term $f_{td}$, independently. 
We compute the baseline errors $\mathrm{MSE}^{td}$ and $\mathrm{MSE}_{\Delta}^{td}$ from this trajectory. 
The normalized metrics are then calculated as:
\begin{equation}\label{eq: nmse_delta}
\mathrm{nMSE} = \frac{\mathrm{MSE}}{\mathrm{MSE}^{td}}, \quad \mathrm{nMSE}_{\Delta} = \frac{\mathrm{MSE}_{\Delta}}{\mathrm{MSE}_{\Delta}^{td}}.
\end{equation}
In practice, we use $\mathrm{nMSE}_{\Delta}$ as the primary criterion for ranking equation candidates. A value of $\mathrm{nMSE}_{\Delta} < 1$ indicates that the inclusion of the candidate spatial evolution term successfully captures interaction-driven dynamics beyond what simple temporal periodicity can explain.

Acknowledge that higher equation complexity can inherently offer greater fitting flexibility and therefore achieve lower fitting errors without necessarily providing a better representation of the underlying spatial interaction process.
To account for this trade-off, we quantify the complexity of spatial interaction equation using the expression length, which is commonly adopted in data-driven equation discovery \citep{guo2025distilling}.
We represent the formula as an expression tree and define complexity as the total number of nodes, counting every occurrence of variables, parameters, constants, and operators.
Because equation complexity can be quantified in different ways, we treat it as an auxiliary indicator for contextualizing the trade-off between fitting accuracy and structural simplicity across candidate equations.

\section{Synthetic experiment}\label{sec: syn}
\begin{figure}[!ht]
    \centering
    \includegraphics[width=0.9\linewidth]{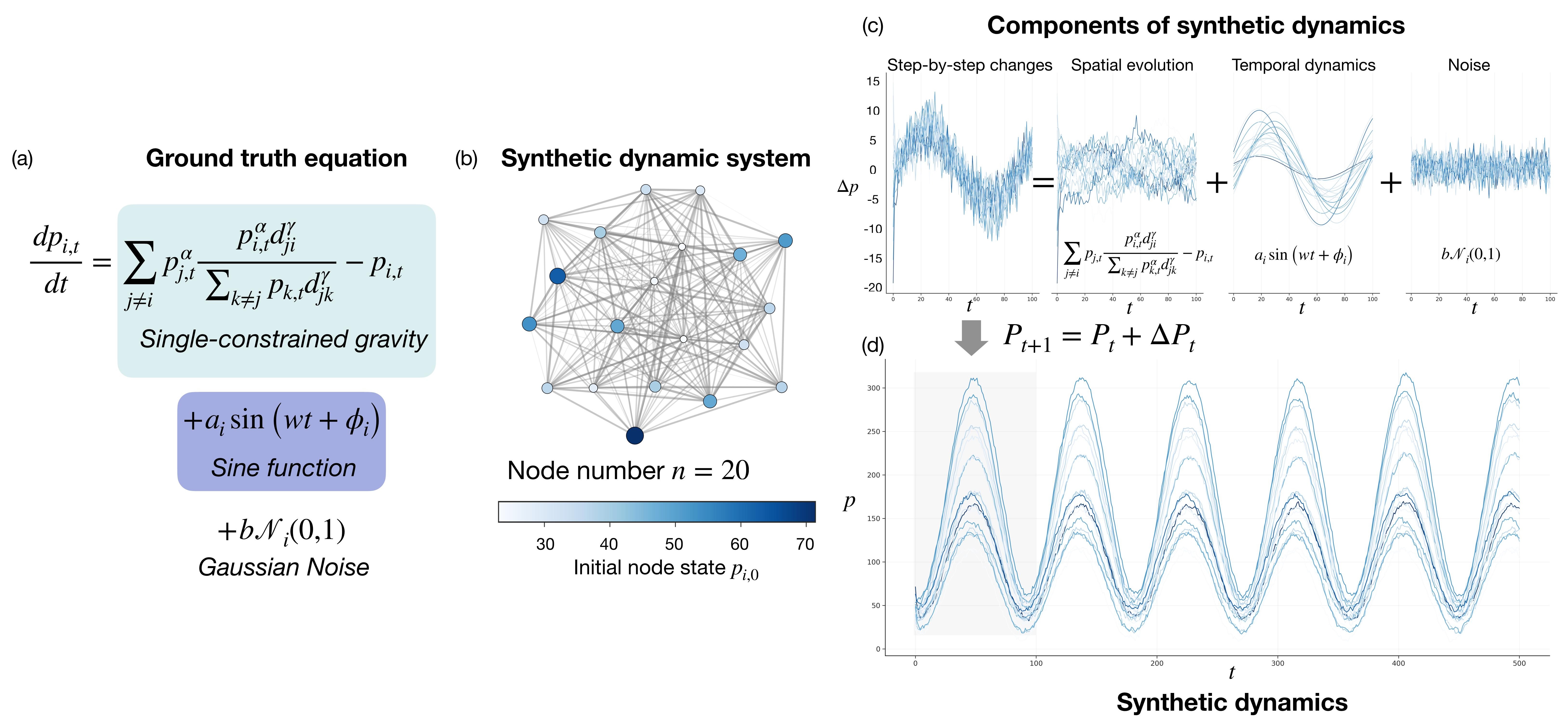}
    \caption{Synthetic spatiotemporal system and dynamics. 
    (a) The ground truth governing equation. 
    (b) A 20-node fully connected spatial network with initial node state $p_{i,0} \sim U(20, 70)$ and edge weight $d_{ij} \sim U(3, 30)$. 
    We colored the lines in (c) and (d) based on the node color in (b) for tracing the dynamics from different initial node states.
    (c) Additive components of the governing equation driving temporal change: a spatial evolution term with the single-constrained spatial gravity as spatial interaction equation, a sine function as uniform periodic forcing, and Gaussian noise. 
    (d) The resulting synthetic time-series dynamics.}
    \label{fig: Synthetic_data}
\end{figure}

In this section, we evaluate the U-Discovery framework using synthetic experiments to assess whether it can reliably fit equation candidates and whether it can correctly identify the true governing spatial interaction equation from a set of candidates. 
As shown in Figure \ref{fig: Synthetic_data} (a), a ground truth governing equation is defined, with the Single-constrained Gravity model \citep{barbosa2018human} as the ground truth spatial interaction equation; a sine function as the temporal dynamics equation with amplitude $a_i$ and phase $\phi_i$ under uniform frequency $\omega$; and additive Gaussian noise with scale $\sigma$. 
A synthetic dynamic system is constructed as a fully connected graph with $N=20$ nodes (Figure \ref{fig: Synthetic_data} (b)), following the definition of the dynamic urban system. 
The edge weights $d_{ij}$ and initial node states $p_{i,0}$ are sampled from two uniform distributions, respectively. 
We simulate a $500$-step time series based on the ground truth equation and graph structure.
By iteratively adding the step-by-step changes from the three components of the ground truth equation (Figure \ref{fig: Synthetic_data} (c)) to the state of the previous time step, we derive the synthetic dynamics observation (Figure \ref{fig: Synthetic_data} (d)).

To test the hypothesis generation capabilities of our framework, we constructed a GraphRAG system with the literature collection introduced in Section \ref{sec: candidates}. 
The equation candidates are proposed by a No Invention mode, which strictly translates the relevant priors to candidates without involving knowledge beyond the literature collection. 
Candidates 6 through 11 in Table \ref{tab:equation_complexity} are equation candidates proposed from the HG step.
Each generated candidate explicitly integrates mechanistic principles from existing studies embedded within the knowledge graph. 
For example, the formulation for Rank-based diffusion (Candidate 7 in Table \ref{tab:equation_complexity}) can be traced back to the universal mobility model introduced by \cite{yan2017universal}. 
This study leverages Zipf’s law, postulating that individuals rank potential destinations such that the probability of visitation is inversely proportional to a location's rank. 
The GraphRAG successfully identified and extracted this principle, combining it with a standard diffusion model to dynamically weight the spatial diffusion flows based on rank probabilities.
We also develop a set of spatial evolution terms (Candidates 2 to 5) based on classic spatial interaction equations, as shown in Table \ref{tab:equation_complexity}.
For the temporal dynamics term, we utilize a k-order Fourier series as the general functional form \citep{chen2008wave}, as shown below:
\begin{equation}
    f_{td}(t;\theta_{td}) = \sum_{k=1}^{K}\left[a_k \cos\left(\frac{2\pi k t}{F}\right) + b_k \sin\left(\frac{2\pi k t}{F}\right)\right],
\end{equation}
where $F$ represents the fundamental period of the rhythm (e.g., 24 hours for daily patterns) and $K$ determines the number of harmonics. $\theta_{td}=\{a_1,b_1,\ldots,a_K,b_K\}$ is the parameter set.

\begin{table}[!ht]
\centering
\renewcommand{\arraystretch}{1.8}
\caption{Selected equation candidates for the synthetic experiment}
\label{tab:equation_complexity}
\resizebox{\linewidth}{!}{
\begin{tabular}{cllc}
\hline
& \textbf{Name}
& \textbf{Equation}
& Complexity \\
\hline

1. & Single-constrained gravity model (POW)
& $\displaystyle
T_{ij}
=
p_{i,t}
\frac{p_{j,t}^{\alpha}d_{ij}^{\gamma}}
{\sum_{k\ne i}p_{k,t}^{\alpha}d_{ik}^{\gamma}}
$
& 18 \\

2. & Single-constrained gravity model (EXP)
& $\displaystyle
T_{ij}
=
p_{i,t}
\frac{p_{j,t}^{\alpha}e^{\gamma d_{ij}}}
{\sum_{k\ne i}p_{k,t}^{\alpha}e^{\gamma d_{ik}}}
$
& 22 \\

3. & Classic gravity model (POW)
& $\displaystyle
T_{ij}
=
\frac{p_{i,t}^{\alpha}p_{j,t}^{\beta}}
{d_{ij}^{\gamma}}
$
& 11 \\

4. & Classic gravity model (EXP)
& $\displaystyle
T_{ij}
=
\frac{p_{i,t}^{\alpha}p_{j,t}^{\beta}}
{e^{\gamma d_{ij}}}
$
& 13 \\

5. & Radiation model
& $\displaystyle
\begin{aligned}
T_{ij}
=
\frac{p_{i,t}p_{j,t}}
{\bigl(p_{i,t}+s_{ij,t}\bigr)
 \bigl(p_{i,t}+p_{j,t}+s_{ij,t}\bigr)},\qquad
s_{ij,t}
=
\sum_{k\in\mathcal V\setminus\{i,j\}}
p_{k,t}\,
\mathbb I\!\left(d_{ik}<d_{ij}\right)
\end{aligned}
$
& 19 \\

6. & Distance-weighted diffusion
& $\displaystyle
T_{ij}
=
d_{ij}^{-\gamma}
\bigl(p_{j,t}-p_{i,t}\bigr)
$
& 7 \\

7. & Rank-based diffusion
& $\displaystyle
T_{ij}
=
\left[
\operatorname{rank}_{i}(d_{ij})^{-\gamma}
+
\operatorname{rank}_{j}(d_{ji})^{-\gamma}
\right]
\bigl(p_{j,t}-p_{i,t}\bigr)
$
& 13 \\

8. & KNN diffusion
& $\displaystyle
T_{ij}
=
\frac{d_{ij}^{-\gamma}}
{\sum_{k\in\mathcal N_K(i)}d_{ik}^{-\gamma}}
\bigl(p_{j,t}-p_{i,t}\bigr),
\qquad j\in\mathcal N_K(i)
$
& 12 \\

9. & Saturation-limited diffusion
& $\displaystyle
T_{ij}
=
\frac{d_{ij}^{-\gamma}}
{\sum_{k\ne i}d_{ik}^{-\gamma}}
\left(
\frac{p_{j,t}}{1+p_{j,t}/\kappa}
-
\frac{p_{i,t}}{1+p_{i,t}/\kappa}
\right)
$
& 24 \\

10. & Accessibility-driven drift
& $\displaystyle
T_{ij}
=
p_{i,t}
\frac{d_{ij}^{-\gamma}}
{\sum_{k\ne i}d_{ik}^{-\gamma}}
\left(
\sum_{k\ne j}d_{jk}^{-\gamma}
-
\sum_{k\ne i}d_{ik}^{-\gamma}
\right)
$
& 20 \\

11. & Local--global mixed diffusion
& $\displaystyle
\begin{aligned}
T_{ij}
={}&
(1-\Gamma)d_{ij}^{-\gamma}
\bigl(p_{j,t}-p_{i,t}\bigr)+
\frac{\Gamma\chi}{n}
\bigl(p_{j,t}-p_{i,t}\bigr)
\end{aligned}
$
& 21 \\
\hline
\end{tabular}
}
\par\smallskip
\begin{minipage}{\linewidth}
\footnotesize\raggedright
Detailed definitions of the notation used in the table are provided in Appendix \ref{app:equation_candidate_notes}.
\end{minipage}
\end{table}

The synthetic dynamics, spanning 500 time steps, was partitioned them into training (60\%), validation (20\%), and testing (20\%) sets. 
To ensure robustness and mitigate the impact of initialization bias, each equation was fitted across 10 independent random seeds; the best-fitting performance for each candidate is summarized in Table \ref{tab:model_comparison_syn}.
The results demonstrate that the ground truth equation, Single-constrained Gravity model (POW), consistently outperforms all competing candidates across every metric. 
The distinction is most pronounced in the differential metric, $MSE_\Delta$ (see Eq. \ref{eq: mse_delta}), where the ground truth model achieves a value of 5.234, a 45.5\% reduction in error compared to the second-best performing model, the Single-constrained Gravity model (EXP), while the differences in $MSE$, $R^2$, and Correlation $Corr.$ are less dramatic.
The second-best candidate, the Single-constrained Gravity model (EXP), has the smallest structural difference from the ground truth. 
Even though the only variation is the choice of the distance decay function (exponential instead of power-law), this minor mathematical substitution results in a substantial performance drop.
The remaining equation candidates perform similarly to one another, forming a performance plateau. Models ranging from the Classic Gravity formulations to various diffusion and radiation mechanisms yield $MSE_\Delta$ values tightly clustered between 9.67 and 9.89. 
This indicates that any deviation from the exact ground truth structure prevents the models from accurately recovering the underlying flux dynamics, further confirming the high identifiability of the true equation within this framework.

\begin{table}[!ht]
\caption{Performance of candidates in the synthetic experiment.}
\label{tab:model_comparison_syn}
\resizebox{\linewidth}{!}{
\begin{tabular}{lccccccc}
\toprule
\textbf{Candidates} 
& \textbf{MSE$_\Delta$} 
& \textbf{R$^2_\Delta$} 
& \textbf{Corr.$_\Delta$} 
& \textbf{MSE} 
& \textbf{R$^2$} 
& \textbf{Corr.} 
& \textbf{Complexity} \\
\midrule
\textbf{Single-constrained Gravity model (POW) }
& \textbf{5.2345} & \textbf{0.7293} & \textbf{0.8083} & \textbf{3.2497} & \textbf{0.9994} & \textbf{0.9994} & 18 \\

Single-constrained Gravity model (EXP) 
& 9.6106 & 0.5030 & 0.6837 & 4.5115 & 0.9992 & 0.9992 & 22 \\

Classic Gravity model (POW) 
& 9.6728 & 0.4997 & 0.6832 & 4.3428 & 0.9992 & 0.9992 & 11 \\

Saturation-limited diffusion
& 9.6912 & 0.4988 & 0.6838 & 4.2980 & 0.9992 & 0.9992 & 24 \\

Classic Gravity model (EXP) 
& 9.7263 & 0.4970 & 0.6824 & 4.3142 & 0.9992 & 0.9992 & 13 \\

Distance-weighted diffusion
& 9.7443 & 0.4960 & 0.6817 & 4.3209 & 0.9992 & 0.9992 & 7 \\

Rank-based diffusion
& 9.7628 & 0.4951 & 0.6816 & 4.2932 & 0.9992 & 0.9992 & 13 \\

Local--global mixed diffusion
& 9.7704 & 0.4947 & 0.6816 & 4.2757 & 0.9992 & 0.9992 & 21 \\

KNN diffusion
& 9.7851 & 0.4939 & 0.6811 & 4.2700 & 0.9992 & 0.9992 & 12 \\

\textit{Temporal dynamics} 
& 9.8226 & 0.4920 & 0.6803 & 4.2350 & 0.9992 & 0.9992 & -- \\

Accessibility-driven drift
& 9.8226 & 0.4920 & 0.6803 & 4.2417 & 0.9992 & 0.9992 & 20 \\

Radiation 
& 9.8851 & 0.4888 & 0.6790 & 4.6677 & 0.9992 & 0.9992 & 19 \\
\bottomrule
\end{tabular}
}
\end{table}

\begin{figure}[!htb]
    \centering
    \includegraphics[width=0.9\linewidth]{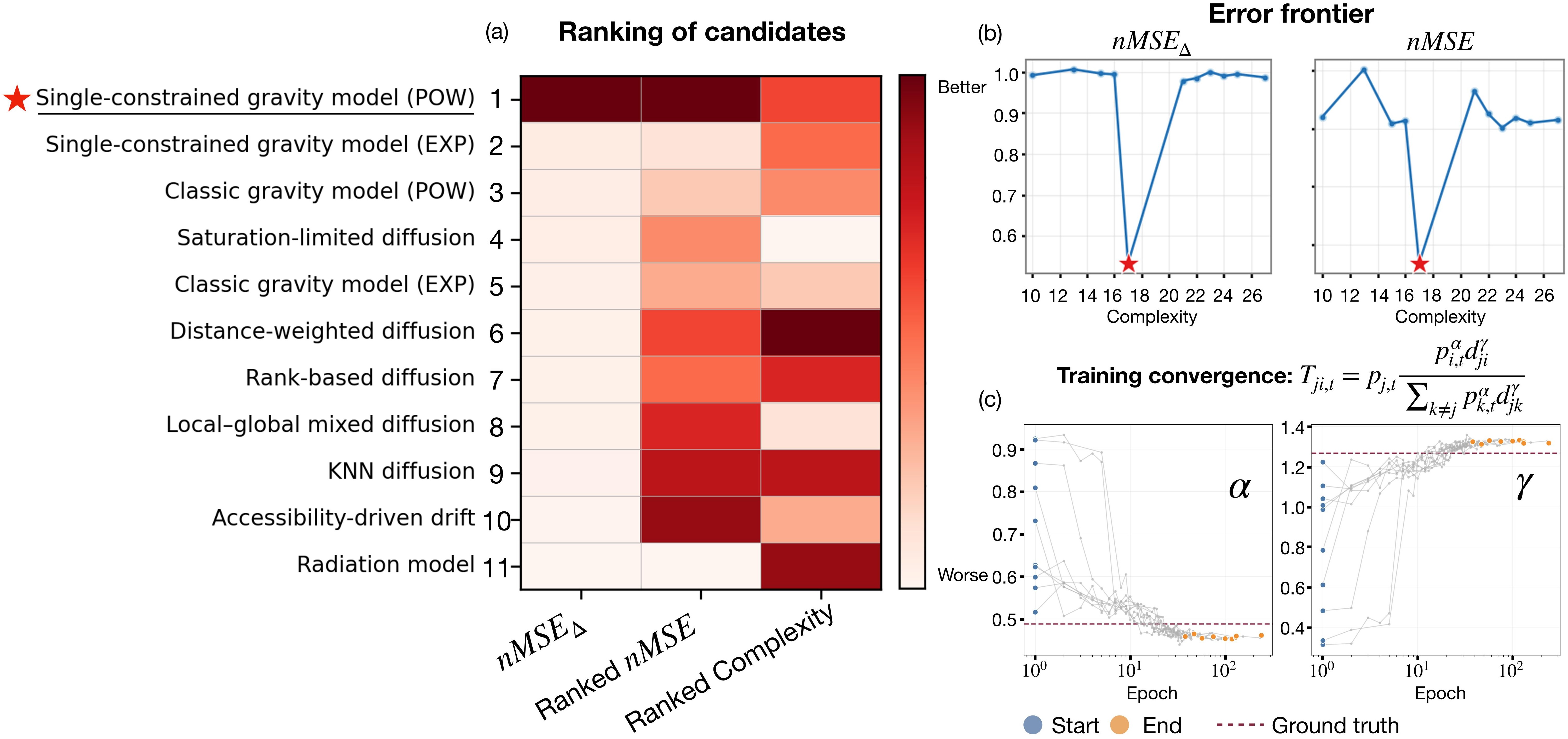}
    \caption{Ranking of equation candidates and parameter convergence.
    (a) Heatmap of equation candidates ranked by $nMSE_\Delta$. 
    The three columns represent $nMSE_\Delta$, the rank of the $nMSE$, and equation complexity. 
    The red star indicates the optimal candidate, the Single-constrained gravity model (POW). 
    The Single-constrained gravity model (POW), marked with a red star, is identified as the optimal candidate.
    (b) Error frontiers showing $nMSE_\Delta$ and $nMSE$ across different complexity. 
    The red stars denote the minimum error achieved by the top-ranked equation.
    (c) Training convergence of the parameters for the ground truth equation during optimization. Blue dots represent the initial parameter values, which converge toward the final estimated values (orange dots) near the target ground truth (red dashed lines).
    }
    \label{fig:syn_results}
\end{figure}

Building on the quantitative results from Table \ref{tab:model_comparison_syn}, we further visualize the model selection process in Figure \ref{fig:syn_results}, where the candidates are ranked by $nMSE_\Delta$ (see Eq. \ref{eq: nmse_delta}). 
We display the raw values of $nMSE_\Delta$, supplemented by the relative ranks of standard $nMSE$ and equation complexity. 
Consistent with the tabular data, the ground truth equation, the Single-constrained Gravity model (POW), is successfully identified as the top-ranked candidate, exhibiting the lowest $nMSE_\Delta$. 
To contextualize this performance against structural simplicity, Figure \ref{fig:syn_results}(b) plots the error frontiers for both $nMSE_\Delta$ and standard $nMSE$. 
By connecting the equations that yield the lowest errors at each discrete level of complexity, we form a Pareto frontier. 
This frontier clearly demonstrates that the top-ranked model achieves the global minimum error, proving that it perfectly captures the underlying dynamics without requiring excessive mathematical complexity. 
Finally, focusing on the ground truth equation, Figure \ref{fig:syn_results}(c) tracks the fitting trajectories of its two core parameters during optimization. 
Across the independent runs using different random seeds, the initial parameter values vary widely but consistently and stably converge toward the true target values. 
Because the synthetic data incorporates inherent noise, which is absorbed during the fitting of the governing equation, we observe a minor but expected deviation between the final estimated parameters and the exact ground truth.

Acknowledge that study of data-driven equation discovery has been increasingly developed for spatial interaction.
However, existing frameworks for discovering spatial interaction equations, such as symbolic regression \citep{cabanas2025human, guo2025distilling}, mainly focus on static scenarios where spatial flows for a given period are observed.
More recent neural symbolic regression, such as Neural Discovery of Network Dynamics in \cite{yu2025discover}, extends equation discovery to dynamic scenarios, but typically relies on additional flow-related information, such as inflow and outflow, to constrain the equation form of interactions.
In contrast, for the dynamic geographic systems considered in this study, the observed data consist of node states and spatial structure, while the spatial interactions are unobserved and all node pairs are treated as potential interactions.
This latent and fully connected interaction setting substantially expands the space of possible equation forms that could explain the observed node-state changes, motivating us to adopt LLM-based hypothesis generation grounded in scientific priors rather than unrestricted symbolic search.
Therefore, these existing frameworks are not directly comparable under the observation setting considered in this study and are not included in our experiments.

\section{Experiments on human mobility data}\label{sec: real exp}
\subsection{Study area and data}\label{sec: data}

\begin{figure}[!htb]
    \centering
    \includegraphics[width=0.9\linewidth]{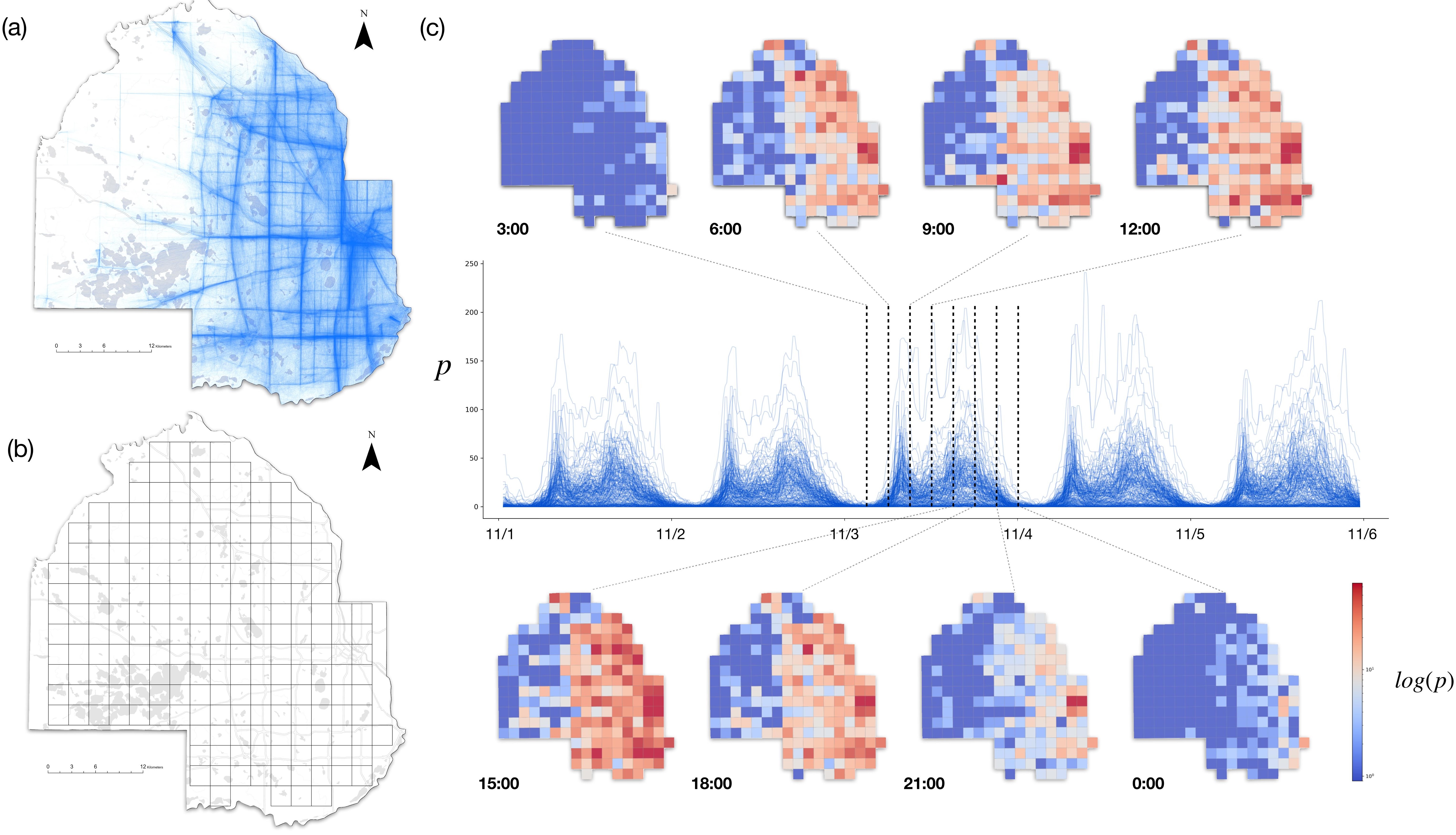}
    \caption{Overview of the mobility dataset and spatiotemporal dynamics used in the empirical experiment. 
    (a) Spatial distribution of the raw trajectory data recorded on November 1, 2021. 
    (b) The study area is partitioned into 2.5 km $\times$ 2.5 km grid cells. 
    (c) Observed urban dynamics aggregated from mobility data showing the time series of active population for each spatial unit. 
    The surrounding maps display snapshots at different time steps over a single day to illustrate changes in the spatial distribution.}
    \label{fig:data}
\end{figure}

In this study, we focus on active population dynamics as an empirical case, with the aim of discovering the spatial interaction equation that best captures mobility-driven population redistribution among spatial units.
Hennepin County is selected as our study area, which is the most populous county in the state of Minnesota in the U.S. 
This county extends from Minneapolis, the urban area with dense commercial and human activity, to the suburbs in the western part of the Twin Cities Metropolitan Area.
To characterize the dynamics of the study area, we extract population fluctuations driven by human mobility using the individual-level mobile positioning dataset from PlaceIQ\footnote{https://www.placeiq.com/}. 
Each record corresponds to a single trip made by an individual, with attributes including the device ID, trip start time, trip duration (in seconds), and the coordinates of the origin and destination locations.
We obtained over 4.76 million trip records from 471{,}323 individuals during the week of November 1--5, 2021.
Figure \ref{fig:data}(a) visualizes the individual-level trip trajectories on November 1, 2021, with each line representing an origin–destination movement within the study area.

Using a 2.5km $\times$ 2.5km grid, we partition the study area into 210 spatial units, as shown in Figure \ref{fig:data}(b), with each unit representing a component of the urban system, and the Euclidean distance between the centroids of units is used to capture the spatial structure. 
Individual trips are then aggregated into consecutive 15-minute intervals to construct time-varying population distributions. 
For each interval, we identify the final stop location of each individual and assign it to the corresponding grid cell. 
The number of individuals whose last observed location falls within a cell is defined as the active population of that spatial unit at that time.
Over the 5-day study period, this discretization yields a total of 480 time steps
By capturing these movement-driven dynamics, spatial interaction between these units is expected to be a fundamental component driving the overall dynamics of the urban system.

\subsection{Results}

We generated 15 additional spatial interaction equations utilizing the partial invention mode, as shown in Table \ref{tab:equation_complexity_2}, and combined them with the equation set in the synthetic experiment to form a total of 26 equation candidates for the empirical experiment. 
Note that the number of candidates evaluated here reflects the experimental setup rather than the proposing capacity of U-Discovery, which can be increased by prompt setting and enriching the knowledge graph with cross-domain literature to broaden the theoretical foundation for equation proposal.
We can categorize the candidates into three distinct modeling paradigms based on their underlying theoretical priors. 
First, topology-enhanced models (e.g., Candidates 2, 3, 11, 14) emphasize the underlying geospatial network, leveraging metrics such as edge and node betweenness to route mobility through structural corridors. 
Second, Spatial Scale and Boundary Models (e.g., Candidates 4, 5, 6, 8, 9, 10, 15) introduce geometric constraints, utilizing piecewise functions, neighborhood thresholds, and community clusters to mathematically distinguish between localized movements and long-range travel. 
Finally, Behavioral and Heuristic Models (e.g., Candidates 1, 7, 12, 13) move beyond pure spatial physics to encapsulate complex human decision-making processes, such as preferential return, exploration probabilities, and rank-based destination choices.
This diverse candidate set demonstrates the LLM's capacity to propose governing equations that are both scientifically reasonable and highly creative in their mathematical synthesis of complex human behavior, facilitating the discovery of spatial interaction mechanisms.
For the temporal dynamics term, we utilize a k-order Fourier series as introduced in Section \ref{sec: syn}.

\begin{table}[!ht]
\renewcommand{\arraystretch}{1.8}
\centering
\caption{Equation candidates proposed under the partial invention mode}
\label{tab:equation_complexity_2}
\resizebox{\linewidth}{!}{
\begin{tabular}{cllc}
\hline
& \textbf{Name}
& \textbf{Equation}
& Complexity \\
\hline

1. & Rank-choice gravity
& $\displaystyle
\begin{aligned}
&T_{ij}
=
p_{i,t}
\frac{p_{j,t}f_r(i,j)}
{\sum_{q\ne i}p_{q,t}f_r(i,q)},\qquad f_r(i,j)
=
\frac{n\alpha}{r_i(j)}+1-\alpha
\end{aligned}
$
& 26 \\

2. & Betweenness-modulated diffusion
& $\displaystyle
T_{ij}
=
(b_i b_j)^{1/2}d_{ij}^{-\eta}
\bigl(p_{j,t}-p_{i,t}\bigr)
$
& 13 \\

3. & Edge-betweenness weighted diffusion
& $\displaystyle
T_{ij}
=
e_{ij}^{\delta}d_{ij}^{-\eta}
\bigl(p_{j,t}-p_{i,t}\bigr)
$
& 11 \\

4. & Cluster-constrained diffusion
& $\displaystyle
T_{ij}
=
\left[
\rho^{\mathrm{in}}
\mathbb{I}_{\{c(i)=c(j)\}}
+
\rho^{\mathrm{out}}
\mathbb{I}_{\{c(i)\ne c(j)\}}
\right]
d_{ij}^{-\eta}
\bigl(p_{j,t}-p_{i,t}\bigr)
$
& 15 \\

5. & Span-adaptive diffusion
& $\displaystyle
\begin{aligned}
T_{ij}
&=
\left(\frac{d_{ij}}{s_{ij}}\right)^{-\eta}
\bigl(p_{j,t}-p_{i,t}\bigr),\qquad
s_{ij}
=
\sqrt{s_i s_j},
\qquad
s_u
=
\frac{1}{k}
\sum_{\ell\in\mathcal N_k(u)}d_{u\ell},
\quad u\in\{i,j\}
\end{aligned}
$
& 23 \\

6. & KNN-limited diffusion
& $\displaystyle
T_{ij}
=
\frac{
\mathbb{I}_{\{j\in\mathcal N_k(i)\}}
+
\mathbb{I}_{\{i\in\mathcal N_k(j)\}}
}{2}
d_{ij}^{-\eta}
\bigl(p_{j,t}-p_{i,t}\bigr)
$
& 13 \\

7. & Explore--return diffusion
& $\displaystyle
\begin{aligned}[t]
&T_{ij}
=
p_{i,t}
\left[
(1-P_{\mathrm{new}})w_{ij}^{\mathrm{return}}
+
P_{\mathrm{new}}w_{ij}^{\mathrm{explore}}
\right], \qquad
P_{\mathrm{new}}
=
\frac{1}{1+\lambda(\ln S+C)},\\
&w_{ij}^{\mathrm{return}}
=
\frac{
\mathbb{I}_{\{j\in\mathcal N(i)\}}d_{ij}^{-1}
}{
\sum_{q\in\mathcal N(i)}d_{iq}^{-1}
}, \qquad
w_{ij}^{\mathrm{explore}}
=
\frac{
\mathbb{I}_{\{j\ne i\}}d_{ij}^{-\eta_e}
}{
\sum_{q\ne i}d_{iq}^{-\eta_e}
}
\end{aligned}
$
& 49 \\

8. & Dual-layer overlap diffusion
& $\displaystyle
\begin{aligned}
T_{ij}
&=
\left[
\Gamma K_{\ell}(d_{ij})
+
(1-\Gamma)K_c(d_{ij})
\right]
\bigl(p_{j,t}-p_{i,t}\bigr),\qquad K_{\ell}(d)
=d^{-\eta_{\ell}},
\qquad
K_c(d)=d^{-\eta_c}
\end{aligned}
$
& 17 \\

9. & Two-scale piecewise diffusion
& $\displaystyle
\begin{aligned}
T_{ij}
&=
K(d_{ij})
\bigl(p_{j,t}-p_{i,t}\bigr), \qquad
K(d)
=
\begin{cases}
a\,d^{-\eta_1}, & d<r_0,\\
b\,d^{-\eta_2}, & d\ge r_0,
\end{cases}
\qquad
b=a\,r_0^{\eta_2-\eta_1}
\end{aligned}
$
& 27 \\

10. & Span- and cluster-aware diffusion
& $\displaystyle
\begin{aligned}[t]
&T_{ij}
=
\left(\frac{d_{ij}}{s_{ij}}\right)^{-\eta}
\left[
\rho^{\mathrm{in}}
\mathbb{I}_{\{c(i)=c(j)\}}
+
\rho^{\mathrm{out}}
\mathbb{I}_{\{c(i)\ne c(j)\}}
\right]
\bigl(p_{j,t}-p_{i,t}\bigr),\\
&s_{ij}
=
\sqrt{s_i s_j},
\qquad
s_u
=
\frac{1}{k}
\sum_{\ell\in\mathcal N_k(u)}d_{u\ell},
\quad u\in\{i,j\}
\end{aligned}
$
& 31 \\

11. & Corridor-enhanced diffusion
& $\displaystyle
T_{ij}
=
\left(
1+\mu\frac{e_{ij}}{\bar e}
\right)
d_{ij}^{-\eta}
\bigl(p_{j,t}-p_{i,t}\bigr)
$
& 15 \\

12. & Rank-truncated + tail choice diffusion
& $\displaystyle
\begin{aligned}
T_{ij}
&=
p_{i,t}
\frac{p_{j,t}w_{ij}}
{\sum_{q\ne i}p_{q,t}w_{iq}}, \qquad
w_{ij}
=
\begin{cases}
\dfrac{\lambda S}{r_i(j)}+1-\lambda,
& r_i(j)\le R,\\[2mm]
c\,d_{ij}^{-\eta},
& r_i(j)>R
\end{cases}
\end{aligned}
$
& 50 \\

13. & State-adaptive diffusion
& $\displaystyle
\begin{aligned}
T_{ij}
=
\left(\beta_r+\beta_t A_{i,t}\right)
w_{ij}
\bigl(p_{j,t}-p_{i,t}\bigr),\qquad
A_{i,t}
=
\sum_{q\ne i}w_{iq}p_{q,t},
\qquad
w_{ij}
=
\frac{d_{ij}^{-\eta}}
{\sum_{q\ne i}d_{iq}^{-\eta}}
\end{aligned}
$
& 28 \\

14. & Betweenness-augmented gravity
& $\displaystyle
T_{ij}
=
\kappa\,
p_{i,t}^{\alpha}p_{j,t}^{\beta}
e_{ij}^{\delta}d_{ij}^{-\gamma}
$
& 17 \\

15. & KNN two-band diffusion
& $\displaystyle
T_{ij}
=
\left[
a\,
\mathbb{I}_{\{
j\in\mathcal N_k(i)\,\lor\,i\in\mathcal N_k(j)
\}}
+
b\,
\mathbb{I}_{\{
j\notin\mathcal N_k(i)\,\land\,i\notin\mathcal N_k(j)
\}}
\right]
d_{ij}^{-\eta}
\bigl(p_{j,t}-p_{i,t}\bigr),
\quad 0\le b\le a
$
& 15 \\
\hline
\end{tabular}
}
\par\smallskip
\begin{minipage}{\linewidth}
\footnotesize\raggedright
Detail explanation of the notation in equation candidates is provided in Appendix \ref{app:equation_candidate_notes}.
\end{minipage}
\end{table}

\begin{figure}[!ht]
    \centering
    \includegraphics[width=0.9\linewidth]{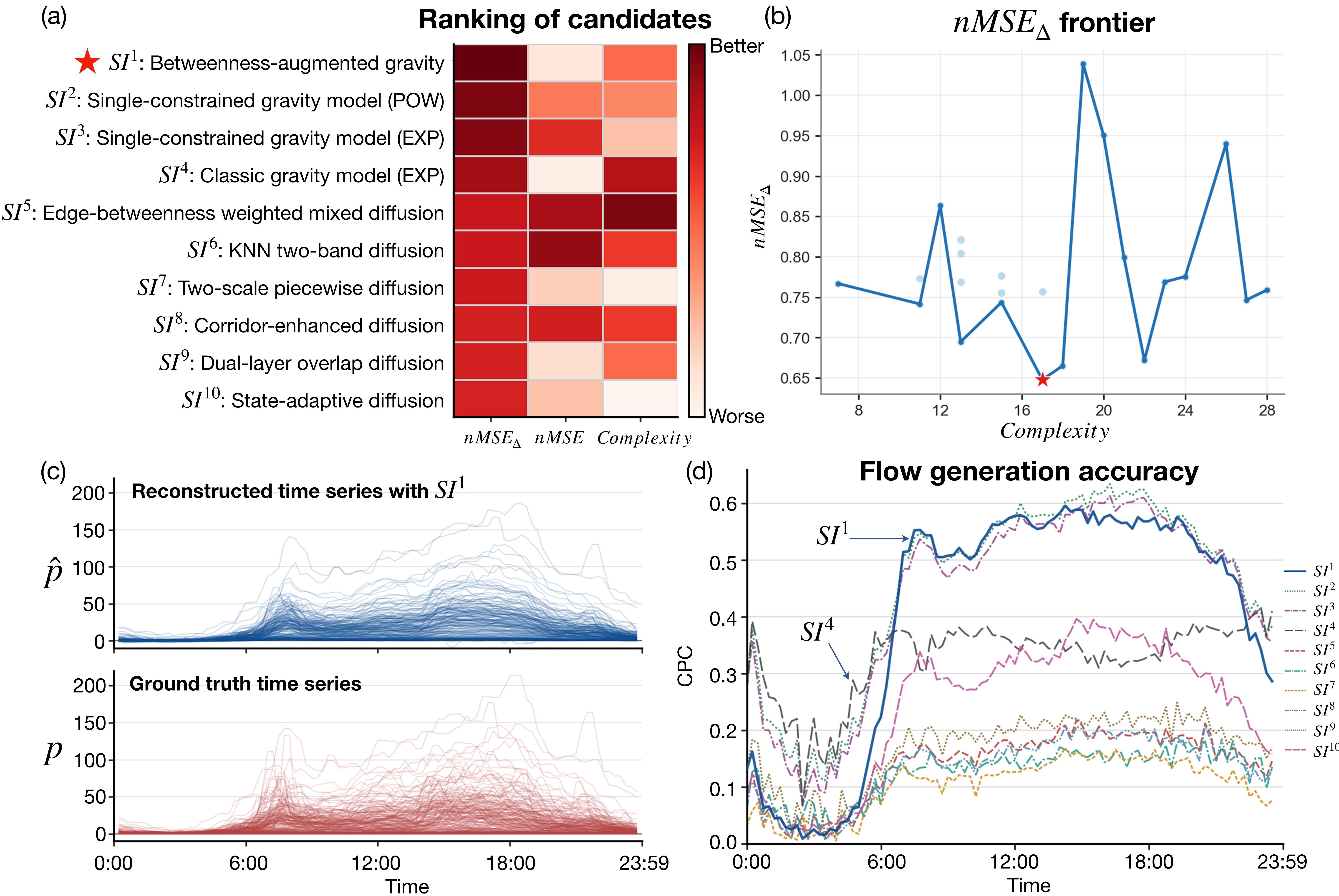}
    \caption{Ranking and validation of spatial interaction equations. 
    (a) Heatmap of candidate performance. Equation candidates are ranked by $nMSE_\Delta$. The three columns display $nMSE_\Delta$ values alongside the relative ranks for standard $nMSE$ and equation complexity. 
    (b) Error Frontiers of $nMSE_\Delta$, with the red star indicating the Betweenness-augmented gravity equation.
    (c) The reconstructed time series with $SI^1$ and the observed time series for all 210 spatial units in the study area.
    (d) The accuracy of generated flow using the top 5 candidates, evaluated by CPC.
    }
    \label{fig:case_results}
\end{figure}

The observed urban dynamics comprise 480 time steps of system snapshots, which are partitioned into a 60\% training set, 20\% validation set, and 20\% testing set.
For each candidate equation, the parameters are calibrated on the training set, and the best-performing checkpoint is selected using the validation set to reduce overfitting to the training period. 
All candidates are calibrated using UrbanDE-Net under identical settings.
Final performance is then evaluated on the testing set. 
To account for spatial heterogeneity in the real-world observations, we adopt the local fitting mode, allowing UrbanDE-Net to estimate node-specific parameter sets for each spatial unit.
The performance of each equation candidate on the testing set is summarized in Table \ref{tab:candidate_models_mse}.
The results reveal a clear trade-off between fitting capacity and structural complexity. 
The highly complex Explore-return diffusion (Complexity $49$) achieves the lowest $nMSE_\Delta$ ($0.6262$) but suffers from severe error accumulation in absolute state prediction ($nMSE = 6.9246$), indicative of overfitting to local gradients. 
In contrast, the Betweenness-augmented gravity equation emerges as the most robust candidate. 
By successfully combining traditional spatial constraints with network topology, it achieves a highly competitive $nMSE_\Delta$ of $0.6479$ and a stable $nMSE$ of $1.1145$ at a moderate complexity of $17$. 
Notably, this proposed formulation outperforms all baseline physics-inspired models, including the Classic Gravity and Radiation models. 
Furthermore, simpler topology-driven equations, such as the Edge-betweenness weighted mixed diffusion (Complexity $11$), demonstrate that integrating correct structural priors can yield strong predictive performance ($nMSE_\Delta = 0.7415$) without the risk of over-parameterization.
In Figure \ref{fig:case_results}(a) and (b), we visualized the results using a heatmap and error frontiers, respectively. 
We discarded the top 10\% most complex equations to prevent distorting the analysis due to over-parameterization and selected the top 10 performing candidates based on their $nMSE_\Delta$ for detailed visual ranking.
As shown in the heatmap (Figure \ref{fig:case_results}(a)), the Betweenness-augmented gravity equation ranks first among all viable candidates, and balances the complexity as shown in the frontier curves, highlighted by the red star (Figure \ref{fig:case_results}(b)).

\begin{table}[!ht]
\centering
\caption{Performance comparison of all equation candidates.}
\label{tab:candidate_models_mse}
\setlength{\tabcolsep}{6pt}
\renewcommand{\arraystretch}{1.15}
\begin{tabular}{p{8.2cm}rrr}
\toprule
\textbf{Equation candidates}
& \textbf{$nMSE_\Delta$}
& \textbf{$nMSE$}
& \textbf{Complexity} \\
\midrule
Explore--return diffusion
& 0.6262 & 6.9246 & 49 \\

Betweenness-augmented gravity
& 0.6479 & 1.1145 & 17 \\

Single-constrained gravity model (POW)
& 0.6650 & 1.0622 & 18 \\

Single-constrained gravity model (EXP)
& 0.6717 & 1.0477 & 22 \\

Classic gravity model (EXP)
& 0.6947 & 1.1148 & 13 \\

Edge-betweenness weighted mixed diffusion
& 0.7415 & 1.0332 & 11 \\

KNN two-band diffusion
& 0.7436 & 1.0145 & 15 \\

Two-scale piecewise diffusion
& 0.7465 & 1.0932 & 27 \\

Corridor-enhanced diffusion
& 0.7551 & 1.0468 & 15 \\

Dual-layer overlap diffusion
& 0.7567 & 1.0998 & 17 \\

State-adaptive diffusion
& 0.7588 & 1.0856 & 28 \\

Span- and cluster-aware diffusion
& 0.7599 & 1.0615 & 31 \\

Span-adaptive diffusion
& 0.7689 & 1.0574 & 23 \\

Rank-based diffusion
& 0.7689 & 1.0748 & 13 \\

Classic gravity model (POW)
& 0.7731 & 1.0819 & 11 \\

Saturation-limited diffusion
& 0.7755 & 1.0404 & 24 \\

Cluster-constrained diffusion
& 0.7769 & 1.0105 & 15 \\

Distance-weighted diffusion
& 0.7800 & 1.0556 & 7 \\

Local--global mixed diffusion
& 0.7991 & 1.0485 & 21 \\

Betweenness-modulated diffusion
& 0.8039 & 1.0552 & 13 \\

KNN-limited diffusion
& 0.8213 & 1.0654 & 13 \\

KNN diffusion
& 0.8634 & 1.0439 & 12 \\

Rank-truncated + tail choice diffusion
& 0.9088 & 0.9850 & 50 \\

Rank-choice gravity
& 0.9397 & 2.1683 & 26 \\

Accessibility-driven drift
& 0.9504 & 1.0089 & 20 \\

Radiation model
& 1.0386 & 1.0585 & 19 \\
\bottomrule
\end{tabular}
\end{table}

To validate the top-performing equation, we evaluated the ability to reconstruct time series using the full governing equation and recover flow using the spatial evolution term. 
In Figure \ref{fig:case_results}(c), we present the reconstructed time series using the governing equation with betweenness-augmented gravity equation ($SI^1$) and the ground truth time series for all 210 spatial units in the study area.
The governing equation with $SI^1$ shows the ability to reproduce the overall temporal trend and the heterogeneous dynamics across spatial units.
However, the reconstruction tends to smooth abrupt peaks and sharp short-term changes, indicating limited ability to capture sudden fluctuations in the observed dynamics.
In Figure \ref{fig:case_results}(d), we compared the generated flows using equation candidates, $\hat T_{ij,t}$, with the observed flows that redistribute population between snapshots, and quantified the similarity between them using the Common Part of Commuters (CPC) \citep{barbosa2018human}.
The performance of these equations is broadly consistent with the \(nMSE_{\Delta}\) ranking: for the top three equations, the betweenness-augmented gravity equation and the power-law and exponential single-constrained gravity equation, also formed the highest-CPC group.
Their CPC values peaked at 0.596, 0.634, and 0.613 and averaged 0.402, 0.454, and 0.436, respectively.
Interestingly, the relative performance of the equations varies across time periods. For example, although $SI^1$ shows outstanding performance during the daytime, the Classic gravity model (EXP) ($SI^4$) outperforms during nighttime and early morning, suggesting that different spatial interaction equations may better characterize the dynamics at different times of day.

\section{Discussion}\label{sec: disscussion}
\subsection{Explore the scale dependence in equation discovery}

\begin{figure}[!ht]
    \centering
    \includegraphics[width=0.9\linewidth]{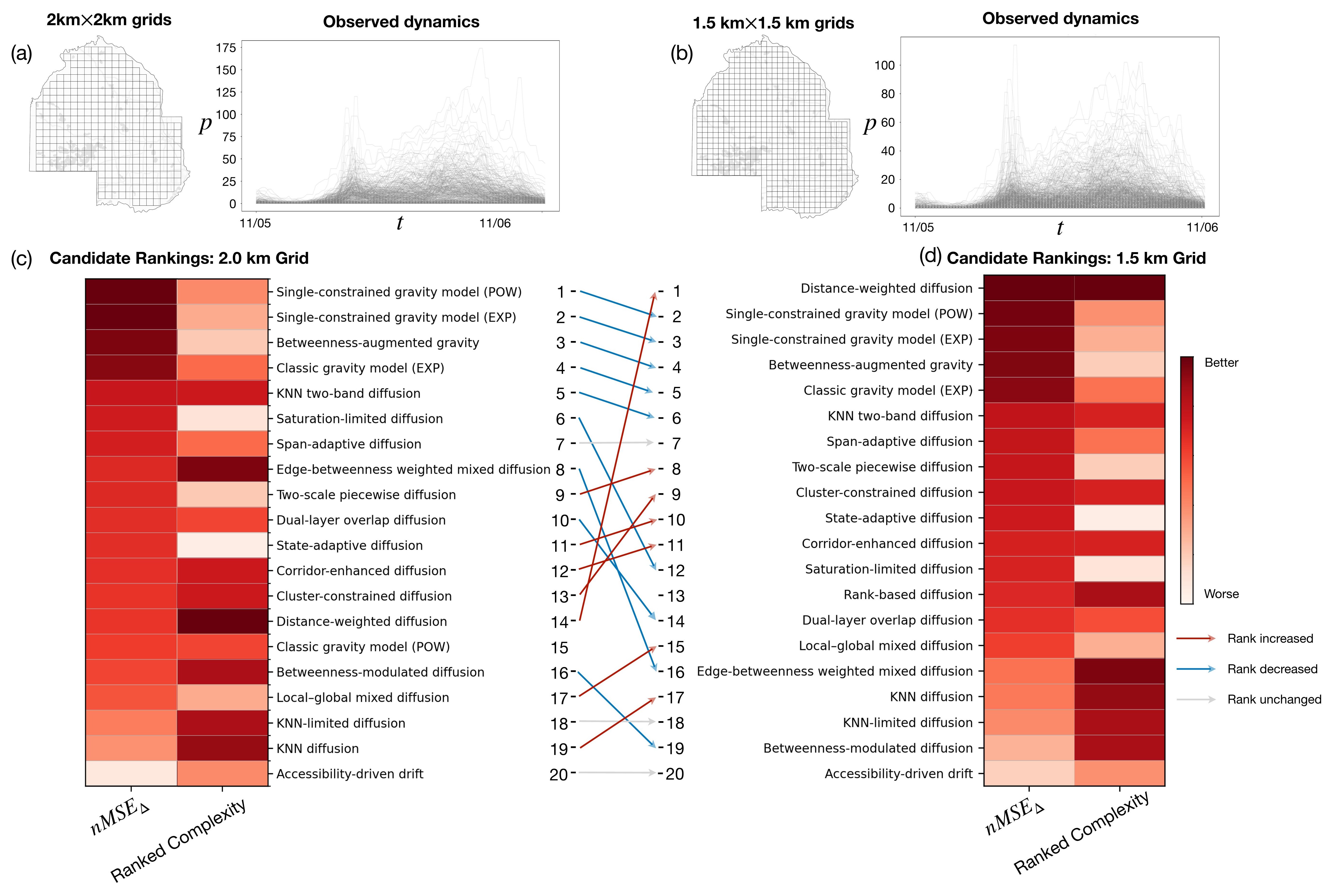}
    \caption{Ranking shifting of equation candidates across spatial scales. 
    Top panels illustrate the spatial grid and corresponding dynamic profiles for (a) the 2.0 km x 2.0 km grids and (b) the 1.5 km x 1.5 km grids. 
    Bottom panels (c) and (d) display heatmaps of candidate rankings at the 2.0 km and 1.5 km grids, respectively, sorted by $nMSE_\Delta$. 
    The connecting arrows between (c) and (d) illustrate rank shifts resulting from the change in scale: red arrows indicate a model improved its relative ranking, blue arrows indicate a drop in ranking, and gray arrows denote no change in rank.}
    \label{fig: diff_scale}
\end{figure}

It is well established in spatial analysis that observed spatial processes and their underlying relationships can vary substantially with the scale of observation, a phenomenon closely associated with scale dependence and the Modifiable Areal Unit Problem (MAUP) \citep{fotheringham1991modifiable}.
Accordingly, the governing spatial interaction equation identified from empirical observations should be interpreted as scale-dependent.
To examine the scale dependence in U-Discovery, we conduct two independent equation discoveries at different spatial scales. 
Specifically, we aggregated the mobility data into 2.0 km $\times$ 2.0 km and 1.5 km $\times$ 1.5 km grids using the same data processing pipeline introduced in Section \ref{sec: data}. 
As U-Discovery is not restricted to a predefined spatial or temporal scale, the same discovery process can be applied consistently across both spatial scales, enabling direct comparison of the discovered equations.
The spatial grids and their aggregated temporal dynamics at different scales are illustrated in Figures \ref{fig: diff_scale}(a) and \ref{fig: diff_scale}(b), respectively.

However, despite the visual similarity in the dynamic data profiles, our equation discovery evaluation reveals that the underlying mechanistic representations are highly sensitive to spatial scale. 
As shown in the performance rankings in Figures \ref{fig: diff_scale}(c) and \ref{fig: diff_scale}(d), changing the spatial scale triggers substantial shifts in model suitability. 
The most prominent change is the significant rise of the Distance-weighted diffusion model, which drastically improved its ranking to become the top-performing candidate at the 1.5 km grids. 
Conversely, models such as the Betweenness-augmented gravity model experienced a drop in their absolute ranking. Notably, while the absolute ranks shifted, the relative hierarchy among the original top five candidate equations remained remarkably stable across both scales. 
In contrast, the equation candidates originally ranked between 6 and 14 exhibited severe volatility, with widespread rank crossing and reordering. 
This suggests that while the dominant spatial interaction equations (the top-tier models) are robust to minor scale modifications, secondary or localized diffusion processes (the mid-tier models) are highly scale-dependent.

\subsection{Limitations and future directions}
The proposed U-Discovery framework still has some limitations.
First, the current U-Discovery framework does not achieve end-to-end autonomous discovery for two reasons.
After proposing equation candidates, the hypothesis generation does not receive empirical feedback from the evaluation phase, which prevents it from completing an iterative refinement of the proposed equation, such as symbolic regression \citep{guo2025distilling, shojaee2024llm}.
Moreover, hypothesis generation is currently focused on the spatial evolution term, while the temporal dynamics term is set as a fixed form with a k-order Fourier series. 
Although Fourier series can flexibly approximate periodic signals \citep{tan2025spatiotemporal}, a fixed harmonic form may inefficiently represent complex, non-stationary urban dynamics and introduce inductive bias into the calibration and identification of the spatial interaction equation.
Future work should therefore establish a closed-loop discovery process in which candidate performance and residual patterns are fed back to the LLM, while the spatial evolution and temporal dynamics terms are jointly proposed, fitted, and refined. Such a coupled process would allow U-Discovery to co-evolve domain-specific functional forms for the complete governing equation, moving the framework toward fully autonomous scientific discovery \citep{li2023autonomous}.

Second, dynamic urban systems exhibit profound complexity that resists simplistic mathematical abstraction. 
Although model-driven approaches, widely used in physical systems, can derive governing equations from first principles, they are difficult to apply in urban contexts due to the absence of first principles and the intrinsic uncertainty of socioeconomic behaviors. 
Our U-Discovery framework addresses this through a data-driven and computationally oriented approach, prioritizing the extraction of empirical patterns over derivation from theoretical principles.
While this is highly effective for identifying plausible spatial interaction equations from observational data, the methodology remains bounded by its data-driven nature.
In particular, the discovery process relies on scientific priors to propose equation candidates rather than directly uncovering the causal mechanism underlying the dynamics \citep{rudy2017data, gao2022autonomous, song2024towards}.
Future advances in network science and complexity science may help alleviate this limitation by providing richer theoretical insights into the generative mechanisms of urban systems. 
Insights from interdisciplinary domains such as statistical physics, ecology, and behavioral science could also provide richer mechanistic priors and expand the space of candidate spatial interaction formulations for future equation discovery.

Third, U-Discovery assumes that, within a specified observation period, the observed system dynamics are governed by a time-invariant equation form, while localized parameters capture spatial heterogeneity across locations. This assumption, however, does not imply that a single immutable equation governs the same urban system under all conditions. 
Even for the same type of observation, the dominant spatial interaction equation may vary across temporal regimes as human behavior, infrastructure conditions, and external disturbances change. 
In our empirical study, the equation discovered from active population dynamics during regular weekdays can be interpreted as reflecting routine population redistribution associated with commuting and daily travel. 
During an emergency event, however, an equation discovered from the same type of active population observations may instead reveal evacuation-driven redistribution shaped by perceived risk, transportation-network disruptions, and shelter accessibility. 
Such context-specific equations could inform urban resilience planning by supporting the estimation of the evacuation flow, thereby the assessment of shelter accessibility and the evaluation of alternative emergency-response strategies.
The relevant equation may also differ across the observed dimensions of the same urban system: transportation dynamics may reveal mechanisms of movement and accessibility, electricity dynamics may reflect spatial coupling in energy supply and demand, and economic dynamics may capture exchanges of labor, goods, or capital. 
Equations discovered from these observations should therefore be interpreted as context-specific representations of distinct urban processes. 
Future research will extend U-Discovery in broader practical applications, from routine active population dynamics to a broader range of urban processes and temporal regimes, examining how context-specific governing equations vary across observation dimensions and event-specific conditions.

\section{Conclusion}\label{sec: conclusion}
In conclusion, this study addresses the long-standing challenge of uncovering the unknown governing equations that drive spatiotemporal urban dynamics, which provide the mathematical formulation to characterize the relationship between flows and changes in urban states. 
Since urban systems lack the deterministic first principles found that commonly exist in physical sciences, the proposed U-Discovery framework provides a novel, data-driven approach to finding governing spatial interaction equations directly from sequential snapshots of spatial distribution data, without requiring the interaction flows to be directly observed.
The framework integrates Large Language Models and GraphRAG to generate scientifically grounded equation candidates (Hypothesis Generation), calibrates these candidates using a neural fitting method called UrbanDE-Net (Neural Fitting Examination), and rigorously ranks them based on a joint criterion of predictive accuracy and mathematical parsimony (Governing Equation Identification). 
Validated through synthetic experiments and empirical dynamics of active population from Hennepin County, Minnesota, U-Discovery successfully identifies optimal, interpretable governing equations that demonstrate its ability to reproduce both the temporal dynamics of active population and the mobility flows driving those dynamics. 
This work bridges the gap between complex-systems theory and empirical observation, providing a robust, scalable foundation for discovering transparent and interpretable governing equations in urban geography and beyond.

As an initial attempt, at the data-driven discovery of spatial interaction equations in urban dynamic studies, U-Discovery demonstrates that the traditional geographical modeling paradigm, which relies on heuristic equation proposal followed by empirical validation, can be significantly augmented by Large Language Models and GeoAI methods. 
We hope that our work, serving as a crucial bridge between spatial big data and complex geographic systems, can empower the next generation of urban digital twins \citep{batty2024digital} and offer fresh perspectives on the critical human-environment challenges confronting modern society \citep{pappalardo2023future}. 
Specifically, uncovering the interpretable candidate equation structures driving various urban dynamics enables more robust simulation and forecasting of large-scale emergencies, including pandemics, wildfires, and extreme weather events. 
Ultimately, these theoretically grounded insights can inform proactive urban resilience planning and infrastructure optimization, thereby mitigating risks and saving lives.

\section*{Acknowledgments}
The authors thank Meicheng Xiong, Sheng Wang, and Jacob Harris for their advice; the editor and the anonymous reviewers for their insightful comments.
\section*{Funding}
This work was supported in part by the Grant-in-Aid of Research, Artistry, and Scholarship program (GIA), University of Minnesota, Twin Cities, in part by the McKnight Land-Grant Professorship Award, University of Minnesota, Twin Cities, and in part by CTS Scholars Award, Center for Transportation Studies, University of Minnesota.
\section*{Conflicts of Interest}
The authors declare no conflicts of interest.
\section*{Data Availability Statement}
The data and code that support the findings of this study are available in GitHub at: \url{https://github.com/GeoDI-Lab/U-Discovery}. 
\bibliographystyle{plainnat}
\bibliography{References}


\appendix
\section{Prompts for scientific prior extraction and equation proposal}\label{app: prompts}

\begin{figure}[!htb]
    \centering
    \includegraphics[width=1\linewidth]{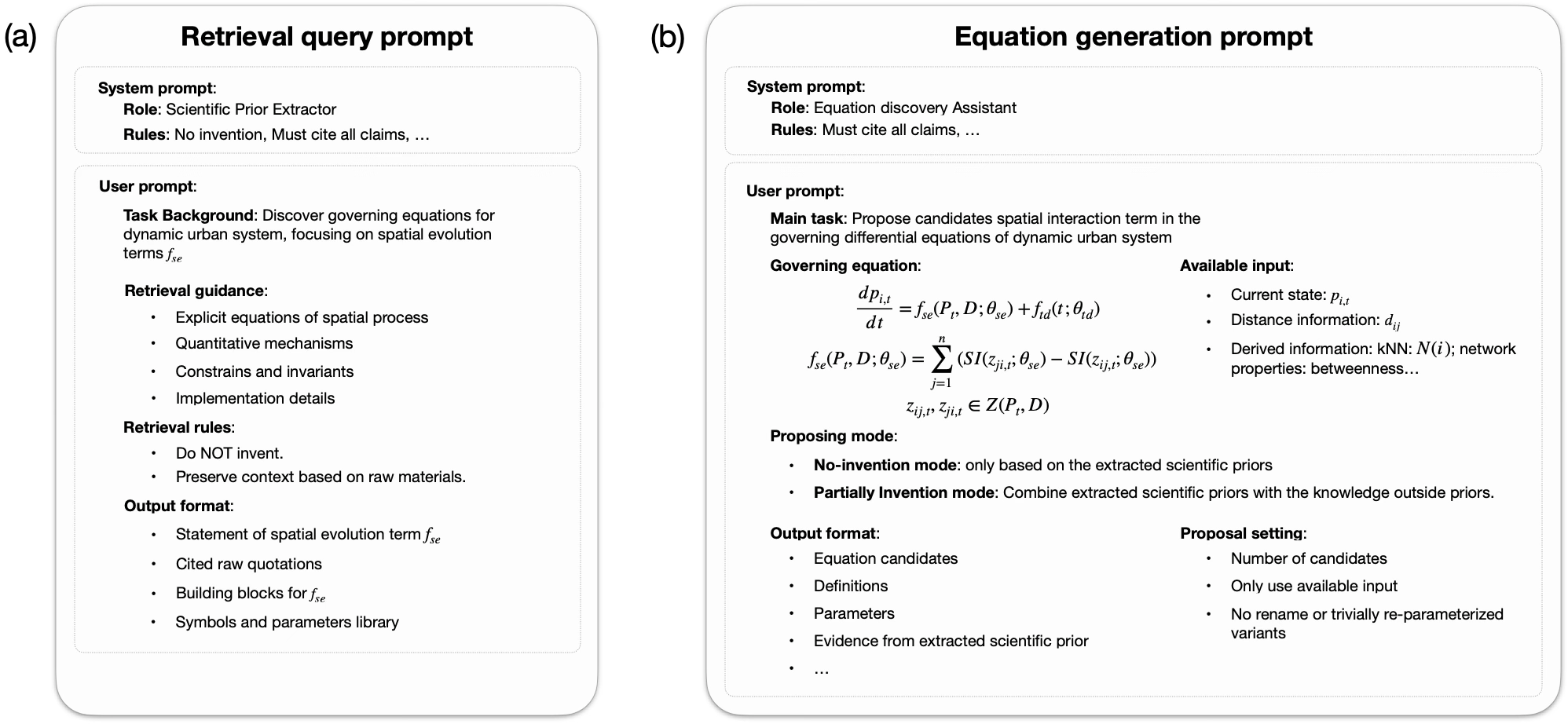}
    \caption{Prompts for (a) scientific prior retrieval and (b) equation candidate generation.}
    \label{fig: prompt}
\end{figure}

The prompts for scientific prior extraction and equation proposal are provided in Figure \ref{fig: prompt} (a) and (b), respectively.
The retrieval query prompt assembles a traceable scientific-prior evidence base for the spatial evolution term \(f_{\mathrm{se}}\). 
Its system prompt assigns an LLM agent the role of a scientific prior extractor, prohibits unsupported invention, and requires citations for all substantive claims. 
The user prompt requests four complementary forms of evidence. Explicit equations provide operator templates that map node states and spatial information to node-level rates of change. 
Quantitative mechanisms explain how processes such as diffusion, flux balance, distance-decay coupling, mobility, and spillovers transmit change across the network. 
Constraints and invariants specify expected mathematical behavior, including conservation, positivity, boundedness, symmetry or directionality, and normalization. 
Implementation details define variables, parameters, and spatial weights and identify potential issues involving identifiability, degeneracy, and numerical stability. 
To retain the meaning of each source, the prompt requires retrieved material to remain close to the original text and preserve its surrounding context. 
The resulting evidence is organized into a source-grounded characterization of \(f_{\mathrm{se}}\), cited excerpts, reusable mathematical building blocks, and a library of supported symbols and parameters.

The equation generation prompt converts the retrieved scientific priors into candidate mathematical forms for \(f_{\mathrm{se}}\). 
It embeds the spatial term within the governing equation \(\mathrm{d}p_{i,t}/\mathrm{d}t=f_{\mathrm{se}}(P_t,D;\theta_{\mathrm{se}})+f_{\mathrm{td}}(t;\theta_{\mathrm{td}})\), thereby separating spatial evolution from the independently modeled temporal component. 
Within this formulation, spatial evolution is represented as the net balance between incoming and outgoing spatial interactions. 
Candidate operators may use the current node states, pairwise distances, and explicitly permitted quantities derived from these inputs, such as distance-based neighborhoods. 
The no-invention mode admits only functional forms directly supported by the extracted priors, whereas the partially inventive mode permits compositions that combine those priors with additional knowledge. 
In both modes, each candidate must provide its mathematical form in both LaTeX and plain text, define any kernels, normalized weights, neighborhood sets, or interaction flows, and specify the meaning and constraints of every parameter. 
It must also identify the required graph inputs and applicable mathematical properties, cite the evidence items supporting its main ingredients, explain its structural difference from the gravity baselines, and disclose any operational assumptions introduced during formulation. 
The proposal settings further control the number of candidates, exclude unavailable inputs, and reject renamed or trivially reparameterized variants, producing a diverse but traceable library of candidate equations.

\section{Notes on the equation candidates}
\label{app:equation_candidate_notes}

This section provides an explanation of the equation candidates shown in Table \ref{tab:equation_complexity} and Table \ref{tab:equation_complexity_2}.
Throughout, \(\mathcal V\) denotes the node set, \(n=|\mathcal V|\), \(p_{i,t}\) is the state of node \(i\) at time \(t\), and \(d_{ij}\) is the distance between nodes \(i\) and \(j\). 
The indicator \(\mathbb I\{A\}\) equals one when \(A\) is true and zero otherwise. All coefficients, exponents, mixing weights, and cutoffs are candidate-specific parameters. Repeated symbols therefore do not imply shared values or meanings across candidates.

In Table~\ref{tab:equation_complexity}For candidates 1--5,
\(T_{ij,t}\) is a directed flow, and$\mathcal S_i(t)=\sum_{j\ne i}\left(T_{ji,t}-T_{ij,t}\right)$.
For candidates 6--11, \(T_{ij,t}\) is a signed pairwise contribution, with $\mathcal S_i(t)=\sum_{j\ne i}T_{ij,t}$.
In Table~\ref{tab:equation_complexity_2}, candidates 1, 7, 12, and 14 use the directed-flow definition, whereas the remaining candidates use the signed-contribution definition. The reverse-direction expression \(T_{ji,t}\) is obtained by exchanging \(i\) and \(j\). Candidates 6, 7, and 11 satisfy \(T_{ji,t}=-T_{ij,t}\) when distances are symmetric; the other candidates are generally direction-dependent. For candidate 7 , \(S\) is the system size, \(C\) is a scalar offset in \(P_{\mathrm{new}}\), and \(\mathcal N(i)\) is the return-neighborhood set. For candidate 12, \(S\) is the system size, \(R\) is the rank cutoff, and \(c\) controls the distance-weighted tail. The functions \(K_{\ell}\) and \(K_c\) in candidate 8 are the two layer-specific distance kernels. Finally,  \(\rho^{\mathrm{in}}\) and \(\rho^{\mathrm{out}}\) are the within- and between-cluster weights, respectively, and \(r_0\) is the distance threshold separating the two spatial scales.

The reported complexity is calculated for the pairwise mechanism only. It excludes the outer aggregation, the net-flow operation, terminal node-balance terms, and any coefficient multiplying the complete spatial term. Parameters appearing within the candidate itself are retained. Auxiliary definitions are recursively expanded at every occurrence. For example, expanding \(s_{ij,t}\) at both occurrences in the radiation
model gives a complexity of \(19\), whereas treating it as a precomputed quantity gives a complexity of \(13\).
The notation \(r_i(j)\), equivalently \(\operatorname{rank}_{i}(d_{ij})\), denotes the distance rank of node \(j\) relative to node \(i\). The sets \(\mathcal N_K(i)\) and \(\mathcal N_k(i)\) denote candidate-specific nearest-neighbor sets of
sizes \(K\) and \(k\), respectively. The quantities \(b_i\) and
\(e_{ij}\) denote node and edge betweenness centrality, \(\bar e\) is the mean edge betweenness over valid edges, and \(c(i)\) is the cluster label of node \(i\). 
\end{document}